\documentclass[aps,prd,amsmath,amssymb,showpacs]{revtex4}

\usepackage{epsfig,float}
\usepackage{graphicx}
\usepackage{dcolumn}
\usepackage{morefloats}
\usepackage{color}
\usepackage{slashed}
\usepackage{bbm}
\usepackage{bm}
\usepackage{mathtools}
\usepackage{booktabs}
\usepackage{multirow}
\usepackage{verbatim}
\usepackage{textcomp,nicefrac}
\allowdisplaybreaks[4]

\newcommand{\la}{\left\langle}
\newcommand{\ra}{\right\rangle}

\newcommand{\half}{\nicefrac{1}{2}}

\begin{document}
\title{Study of heavy quark conserving weak decays in the quark model}

\author{Peng-Yu Niu$^{1,2,3,4}$\footnote{E-mail: niupy@m.scnu.edu.cn}, Qian Wang$^{2,3}$\footnote{E-mail: qianwang@m.scnu.edu.cn}, Qiang Zhao$^{1,4}$~\footnote{E-mail: zhaoq@ihep.ac.cn} }

\affiliation{ 1) Institute of High Energy Physics, Chinese Academy of Sciences, Beijing 100049, P.R. China}
\affiliation{ 2) Guangdong Provincial Key Laboratory of Nuclear Science, Institute of Quantum Matter, South China Normal University, Guangzhou 510006, China }
\affiliation{ 3) Guangdong-Hong Kong Joint Laboratory of Quantum Matter, Southern Nuclear Science Computing Center, South China Normal University, Guangzhou 510006, China }
\affiliation{ 4) University of Chinese Academy of Sciences, Beijing 100049, P.R. China}

\begin{abstract}
Within the framework of a constituent quark model, we investigate the heavy quark conserving weak decays of $\Xi_c$ and $\Xi_b$. This process involves elementary transitions of  $us\to du$ or $s\to ud \bar u$ with the heavy quark as a spectator, while for the charmed baryon in the initial state it can also occur via $cs\to dc$ to conserve the heavy quark flavor in the weak decays. It shows that the heavy quark symmetry (HQS) plays an essential role in the processes where the heavy quark acts as a spectator. For the $\Xi_c$ decays, the dominance of the pole terms in the parity-conserving transitions is evident. This can naturally explain the sizeable branching ratio for $\Xi_c^0\to\Lambda_c\pi^-$ as recently reported by LHCb. The same mechanism predicts a large branching ratio for $\Xi_c^+\to\Lambda_c\pi^0$.  In contrast, the decays of $\Xi_b$ will only go through the parity-violating processes because of the HQS. In addition, the branching ratios of the $\Xi_c^+\to \Lambda_c \pi^0$ and $\Xi_b^0\to \Lambda_b \pi^0$ processes are also predicted for the further experimental measurement.

\end{abstract}

\date{\today}
%
\maketitle
\section{Introduction}

The heavy quark conserving weak decay (HQCWD) of a heavy baryon is referred to such a process where the heavy flavor quark in the initial baryon survives in the weak decay no matter it has participated the interactions or not. This is an interesting phenomenon which is different from most weak decay processes. For the latter ones most of the observed weak decay modes of the initial ground state heavy baryons, i.e. either the charmed or bottomed baryons, are via the heavy flavor decays into light flavor hadrons. Recently, the branching fraction of $\Xi_c^0\to \Lambda_c \pi^-$ was measured by LHCb~\cite{Aaij:2020wtg} for the first time, with a value of $(0.55\pm0.02\pm0.18)\%$. In the bottom sector the first measurement of the relative rate of $\Xi_b^-\to \Lambda_b \pi^-$ was also done by LHCb~\cite{Aaij:2015yoy}, 
\begin{align}
\frac{f_{\Xi_{b}^{-}}}{f_{\Lambda_{b}^{0}}} \mathcal{B}\left(\Xi_{b}^{-} \rightarrow \Lambda_{b}^{0} \pi^{-}\right)=\left(5.7 \pm 1.8_{-0.9}^{+0.8}\right) \times 10^{-4},
\end{align}
where $f_{\Xi_b^-}$ and $f_{\Lambda_b^0}$ are the fragmentation fractions for $b\to \Xi^-_b$ and $b\to \Lambda_b^0$, respectively. The ratio $f_{\Xi_b^-}/f_{\Lambda_b^0}$ takes a value between 0.1 and 0.3, which leads to a range for  the branching fraction $\mathcal{B}\left(\Xi_{b}^{-} \rightarrow \Lambda_{b}^{0} \pi^{-}\right)$ between $(0.19\pm0.07)\%$ and $(0.57\pm 0.21)\%$. Although $\Xi_c^0\to \Lambda_c \pi^-$ and $\Xi_b^-\to \Lambda_b \pi^-$ are both heavy quark conserved process, their decay mechanisms are not exactly the same. The main difference is that in the $\Xi_b^-$ decay the $b$ quark is a spectator while in the $\Xi_c$ decay the charm quark can be either a spectator or directly  participate the weak interaction via $cs\to dc$. These two processes make it an excellent situation to investigate the light quark correlations.  In particular, the effects of the heavy quark symmetry in these two processes will provide important information about the heavy flavor baryons.

For the theoretical aspects, the HQCWD was first noted by Ref.~\cite{Cheng:1992ff}. With the implementation of the heavy quark symmetry and chiral symmetry it shows that only the transitions between the anti-triplet states are allowed in the heavy quark limit~\cite{Cheng:1992ff,Cheng:2015ckx}. In the case of the $\Xi_b\to \Lambda_b \pi$ decay since the $b$ quark is a spectator the interactions in the light quark sector will only involve an $S$-wave contribution~\cite{Gronau:2015jgh,Li:2014ada}.
So the decay of $\Xi_b\to \Lambda_b \pi$ can be related to the $S$-wave decay of $\Lambda$, $\Sigma$ and $\Xi$ by the Dolen-Horn-Schmid duality~\cite{Dolen:1967zz,Dolen:1967jr}. This turns out to be a useful treatment of the parity-violating contributions~\cite{Gronau:2015jgh}. In addition, the current algebra and soft pion limit were also widely employed to study the HQCWD process~\cite{Sinha:1999tc,Faller:2015oma,Gronau:2016xiq,Voloshin:2000et,Voloshin:2019ngb}. The relevant matrix elements were either calculated by different models, such as the MIT bag model and diquark model, or estimated by making an analogue to the hyperon hadronic weak decay (or meson weak decay) processes. 

With the availability of experimental data for $\Xi_c^0\to \Lambda_c \pi^-$~\cite{Aaij:2020wtg} and $\Xi_b^-\to \Lambda_b \pi^-$\cite{Aaij:2015yoy} from LHCb, one notices that the measured branching ratio for $\Xi_b^-\to \Lambda_b \pi^-$ is the same order of magnitude as most of the theoretical predictions. However, there exist significant discrepancies between the theoretical predictions and experimental measurement for the decay of $\Xi_c^0\to \Lambda_c \pi^-$. This is an indication that the transition mechanism for the charm sector is different from that for the bottom one. This motivates us to carry out a systematic analysis of these HQCWD processes. 

As mentioned earlier, in the HQCWD of $\Xi_b$ the $b$ quark is a spectator while in the decay of $\Xi_c$ the charm quark can actually participate the weak interactions. It suggests that the non-factorizable mechanisms should play a significant role in the charm sector. As the heavy quark flavor is conserved here, the non-relativistic constituent quark model (NRCQM)~\cite{Copley:1979wj,Isgur:1978xj} can be applied to the description of the non-factorizable mechanisms including color suppressed transitions and intermediate state excitations. For the singly heavy baryons, the excitations of the light and heavy degrees of freedom can be well separated. Therefore, it is possible to study the contributions from different excitation modes in observables. This provides a unique tool for probing the quark structures inside these  singly heavy baryon states.

As follows, we first introduce the non-relativistic quark model formalism in Section~\ref{framework}. The numerical results and discussions are presented in Section~\ref{results} and a brief summary is given in Section~\ref{summary}. The quark model  wave functions and some detailed calculations of the transition amplitudes are supplied in Appendix~\ref{app:wavefunction}-\ref{app:amplitudes}.

\section {Framework}
\label{framework}
\subsection{The transition mechanism}

\begin{figure}[ht]
\begin{center}
\includegraphics[scale=0.7]{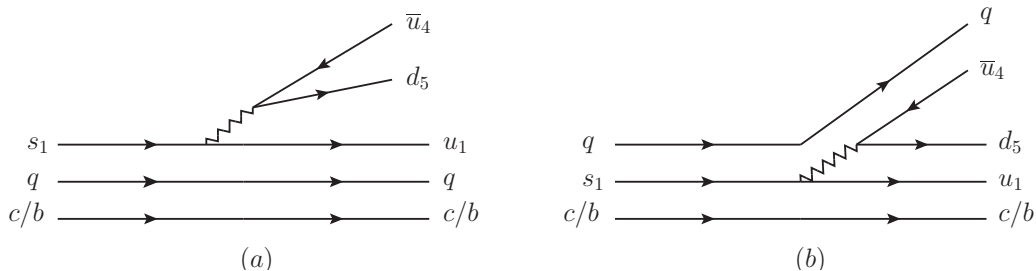}
\caption{Illustrations for the direct weak emission of pion process of $\Xi_Q \to \Lambda_Q\pi(Q=c,b)$ at the quark level. (a) is the direct pion emission(DPE) process and (b) is the color suppressed(CS) process. The subscripts are used to label the quarks which is convenient for constructing the wave functions of hadrons and the effective Hamiltonian.}
\label{fig:DPECS}
\end{center}
\end{figure}

\begin{figure}[ht]
\begin{center}
\includegraphics[scale=0.7]{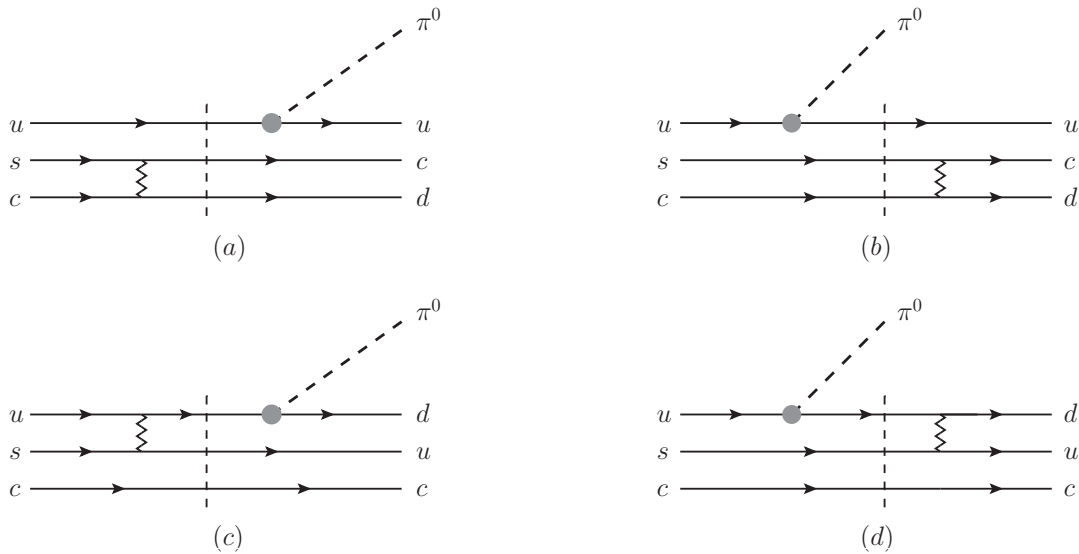}
\caption{Illustrations for the pole terms of $\Xi_c^+\to \Lambda_c \pi^0$ at the quark level. The solid dot stands for the quark-meson interaction.}
\label{fig:PT1}
\end{center}
\end{figure}

\begin{figure}[ht]
\begin{center}
\includegraphics[scale=0.7]{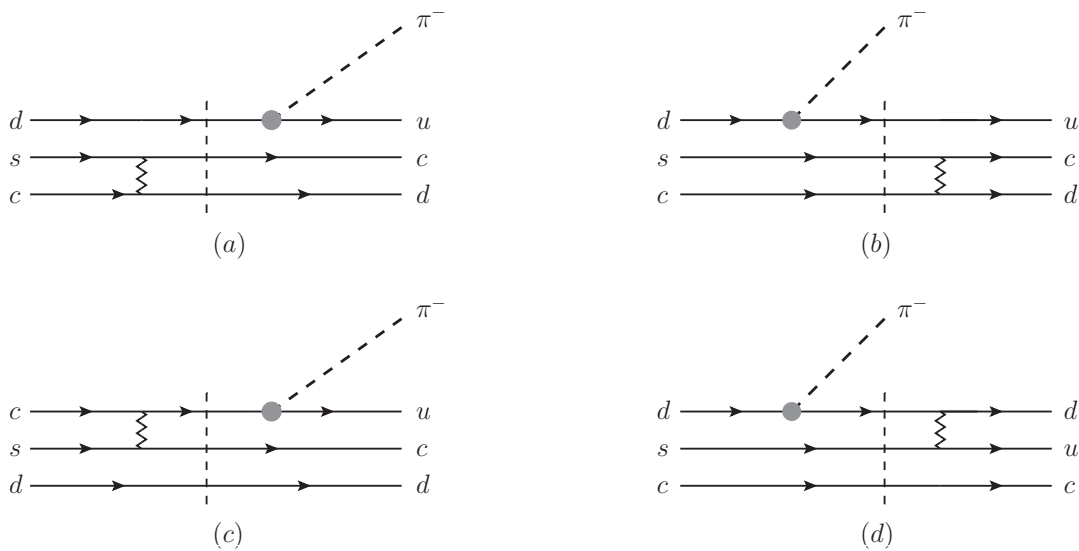}
\caption{Illustrations for the pole terms of $\Xi_c^0\to \Lambda_c \pi^-$ at the quark level. The weak transiton is $sc\to cd $ for the (a) and (b) processes, while the weak transition is $ud \to su $ for the (d) process. The solid dot also stands for the quark-meson interaction.}
\label{fig:PT2}
\end{center}
\end{figure}

\begin{figure}[ht]
\begin{center}
\includegraphics[scale=0.7]{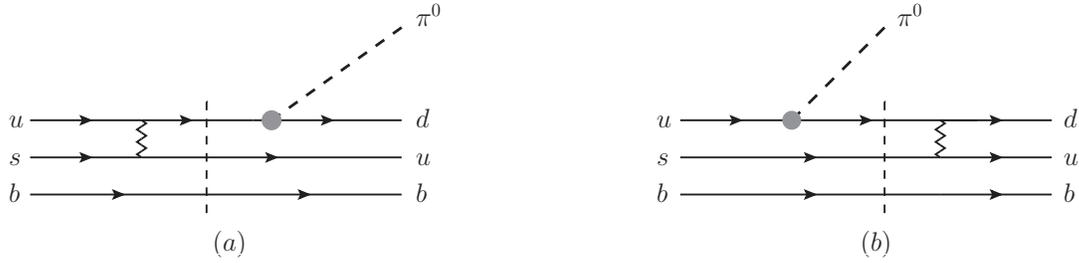}
\caption{Illustrations for the pole terms of $\Xi_b^0\to \Lambda_b \pi^0$  at the quark level. It clearly shows that the $b$ quark acts as a spectator.}
\label{fig:PT3}
\end{center}
\end{figure}

\begin{figure}[ht]
\begin{center}
\includegraphics[scale=0.7]{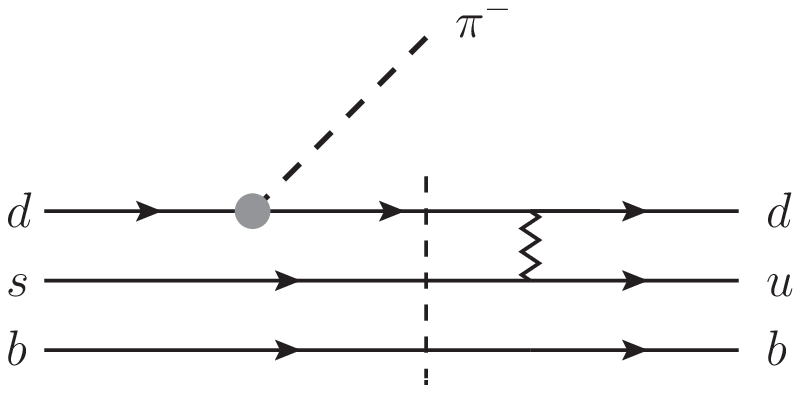}
\caption{Only the B-type pole terms are alllowed for $\Xi_b^-\to \Lambda_b \pi^-$ at the quark level.}
\label{fig:PT4}
\end{center}
\end{figure}

In this work we focus on the decays, $\Xi_c^+ \to \Lambda_c \pi^0$, $\Xi_c^0 \to \Lambda_c \pi^-$, $\Xi_b^- \to \Lambda_b \pi^-$ and $\Xi_b^0 \to \Lambda_b \pi^0$, which are singly-Cabbibo-favored processes. Typical transitions are illustrated in Figs.~\ref{fig:DPECS}-\ref{fig:PT4} which can be categorized as follows: (i) direct pion emission (DPE) (Fig.~\ref{fig:DPECS} (a)); (ii) color suppressed (CS) pion emission (Fig.~\ref{fig:DPECS} (b)); (iii) pole terms (see e.g. Figs.~\ref{fig:PT1}-\ref{fig:PT4}). The first two kinds of transitions involve the elementary process $s\to u d \bar u$. Thus, the heavy quark behaves as a spectator. In contrast, the pole terms describe the non-local feature between the internal-conversion weak transitions and the strong pion emissions, where the heavy quark can be either a spectator (Figs.~\ref{fig:PT1} (c) and (d)) or participate in the weak interaction (Figs.~\ref{fig:PT1} (a) and (b)). Note that the $b$ quark always behaves as a spectator in $\Xi_b \to \Lambda_b \pi$ as shown by Fig.~\ref{fig:PT3} and \ref{fig:PT4}. For the processes that the $b$ quark participates the weak interactions, the amplitudes will be highly Cabbibo suppressed. In Figs.~\ref{fig:PT1}-\ref{fig:PT4} only the singly-Cabbibo-favored processes are illustrated. 

Following the treatment of Ref.~\cite{Niu:2020gjw}, we adopt the chiral quark model~\cite{Manohar:1983md} for the pseudoscalar meson production at the strong interaction vertices~\cite{Li:1997gd,Zhao:2002id,Zhong:2007gp}. Namely, the pion is treated as a fundamental particle which couples to the light quarks via a chiral Lagrangian. For convenience, the processes of Figs.~\ref{fig:PT1} (a) and (c) are all labeled as A-type and the processes of (b) and (d) are labeled as B-type. Similar categorization is adopted for Figs.~\ref{fig:PT2}-\ref{fig:PT4}.

Compared with the DPE and CS processes, the behavior of pole terms should be more complicated because of the contributions from intermediate states. The intermediate states could be any states as long as the quantum number allows, i.e. with $J^P=1/2^+$ or $1/2^-$. In this work, similar to the treatment of the hadronic weak decay processes~\cite{Niu:2020gjw}, only the ground states with quantum number $1/2^+$ and the first orbital states with quantum number $1/2^-$ are considered. 

\subsection{Non-relativistic formalisms }
In this work, we take the non-relativistic form of the effective Hamiltonian which was presented in Refs.~\cite{LeYaouanc:1978ef,LeYaouanc:1988fx}. It was recently applied to hyperon and charmed baryon weak decays in Refs.~\cite{Richard:2016hac,Niu:2020gjw} to confront the progress of experimental measurements. In this section we briefly illustrate the effective Hamiltonian of the weak interaction and the quark-meson coupling of the chiral quark model. 
Before giving the exact form of these operators, it is necessary to emphasize that the heavy quarks $c$ and $b$ are labeled as the third constituent quark in the initial baryon in this work. The corresponding total wave functions of the singly heavy baryons are given in Appendix~\ref{app:wavefunction}.

\subsubsection{Operators of the weak interaction}
For the weak interaction, the four-fermion interaction is employed and it is expressed as~\cite{LeYaouanc:1978ef,LeYaouanc:1988fx,Richard:2016hac}:
\begin{equation}
H_W=\frac{G_F}{\sqrt 2}\int d \bm x \frac12 \{ J^{-,\mu}(\bm x),J^{+}_{\mu}(\bm x) \},
\end{equation}
where
\begin{equation}
\begin{aligned}
J^{+,\mu}(\bm x)&=
\begin{pmatrix}\bar u&\bar c \end{pmatrix}
\gamma^\mu(1-\gamma_5)
\begin{pmatrix}\cos \theta_C & \sin \theta_C \\ -\sin \theta_C &\cos \theta_C \end{pmatrix} \begin{pmatrix} d\\s \end{pmatrix}, \\
J^{-,\mu}(\bm x)&=
\begin{pmatrix}\bar d &\bar s \end{pmatrix} \begin{pmatrix}\cos \theta_C & -\sin \theta_C \\ \sin \theta_C &\cos \theta_C \end{pmatrix} \gamma^\mu(1-\gamma_5)
\begin{pmatrix} u\\c \end{pmatrix}.
\end{aligned}
\end{equation}
This Hamiltonian contains the tree-level operators used for four-fermion interaction processes and it can be always separated into parity-conserving and the parity-violating parts which are labeled with superscript $PC$ and $PV$, respectively.

In this work, only the non-relativistic operators for the $1\to 3$ emission and the $2\to 2$ internal conversion processes need to be considered. For the DPE and CS processes the quark transition process is $s\to u d\bar u$ which is illustrated by Fig.~\ref{fig:DPECS}. With the explicit quark labels given in Fig.~\ref{fig:DPECS}, the corresponding operator can be written as:
\begin{equation}
\begin{aligned}
H_{W,1\to 3}^{PC}
={}&\frac{G_F}{\sqrt2} V_{su}V_{ud} \frac{\beta}{(2\pi)^3}\delta^3(\bm p_1 -\bm p_1'-\bm p_4-\bm p_5 ) \left\{ \la s_1'|I|s_1\ra \la s_5 \bar s_4|\bm \sigma|0\ra \left( \frac{\bm p_5}{2m_5}+ \frac{\bm p_4}{2m_4}\right) \right.\\
&{}-\left[ \left( \frac{\bm p_1'}{2m_1'}+\frac{\bm p_1}{2m_1} \right)\la s_1'|I|s_1\ra -i \la s_1'|\bm \sigma |s_1\ra \times \left( \frac{\bm p_1}{2m_1}-\frac{\bm p_1'}{2m_1'} \right) \right] \la s_5 \bar s_4|\bm \sigma|0\ra  \\
&{} -\la s_1'|\bm \sigma |s_1\ra \left[\left( \frac{\bm p_5}{2m_5}+\frac{\bm p_4}{2m_4} \right) \la s_5 \bar s_4 |I|0\ra
-i  \la s_5 \bar s_4 |\bm \sigma|0\ra\times \left( \frac{\bm p_4}{2m_4}-\frac{\bm p_5}{2m_5} \right)\right] \\
&{}+\la s_1'|\bm \sigma |s_1\ra  \left( \frac{\bm p_1'}{2m_1'}+\frac{\bm p_1}{2m_1} \right) \la s_5 \bar s_4 |I|0\ra \biggr\} \hat\tau^{(+)}_1 \hat I'_\pi,\\
H_{W,1\to 3}^{PV}
={}&\frac{G_F}{\sqrt2}V_{su} V_{ud} \frac{\beta}{(2\pi)^3}\delta^3(\bm p_1-\bm p_1'-\bm p_4-\bm p_5 )
\left( -\la s_1'|I|s_1\ra \la s_5 \bar s_4|I|0\ra + \la s_1'|\bm \sigma|s_1\ra \la s_5 \bar s_4|\bm \sigma|0\ra \right) \hat\tau^{(+)}_1 \hat I'_\pi ,
\end{aligned}
\label{eq:HW13}
\end{equation}
where $\beta$ is the symmetry factor which equals to $2$ for the DPE process and $2/3$ for the CS process; $I$ is the dimension-two unit matrix; $\hat\tau^{(+)}$ is the flavor operator which transforms $s$ to $u$, and $\hat I'_\pi$ is the isospin operator for the pion production process. It has the following form:
\begin{align}
\hat I'_\pi=\begin{dcases}
b^\dag_u b_u                    &\text{for}~ \pi^+,\\
-\frac{1}{\sqrt 2}b^\dag_u b_d   &\text{for}~ \pi^0,\\
\end{dcases}
\end{align}
and will act on the $i$-th light quark of the initial baryon after considering the pion flavor wave function. 
 $\bar s_4$ in Eq.~(\ref{eq:HW13}) stands for the spin of particle $4$ which is an anti-quark. So the particle-hole conjugation~\cite{Racah:1942gsc} should be employed in order to evaluate the spin matrix element containing an anti-quark: 
\begin{align}
\langle j,-m| \to(-1)^{j+m}|j,m\rangle.
\end{align}

\begin{figure}[ht]
\begin{center}
\includegraphics[scale=0.7]{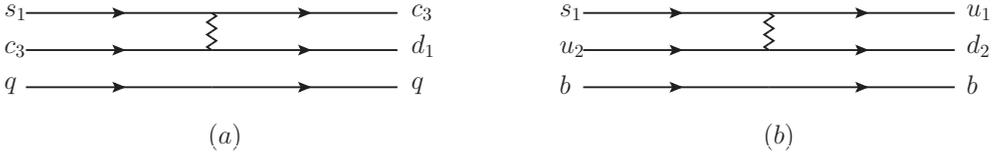}
\caption{The Cabbibo-suppressed $2\to 2$ quark transition process. The figure (a) clearly shows that $c$ quark does not behave as a spectator, while, as shown by figure (b), $b$ is free from the weak interaction. The quark are explicitly labeled by subscripts $1,2$ and $3$ which makes it easier for the expression of the non-relativistic operators.}
\label{fig:w2to2}
\end{center}
\end{figure}

For the pole terms, there are two types of  $2\to 2$ weak transition operators contributing in the amplitudes, i.e. $sc\to cd$ and $su \to ud$. Generally, the transition operator can be written as:
\begin{equation}
H_{W,2\to 2}=\frac{G_F}{\sqrt2}V_1 V_2 \frac{1}{(2\pi)^3} \sum_{i\neq j}\delta^3(\bm p'_i+\bm p'_j-\bm p_i-\bm p_j) \bar u(\bm p_i')\gamma_\mu(1-\gamma_5)u(\bm p_i)\bar u(\bm p_j')\gamma^\mu(1-\gamma_5)u(\bm p_j),
\end{equation}
where $V_1$ and $V_2$ are the Cabbibo-Kobayashi-Maskawa (CKM) matrix elements; $V_1 V_2$ equals to $V_{cs}V_{cs}$ or $V_{us}V_{ud}$ for $\Xi_c\to \Lambda_c \pi$ or $\Xi_b \to \Lambda_b\pi$, respectively. The subscripts $i$ and $j$ are used to indicate the quarks which are acted on by the operators. Considering that the heavy quarks $c$ or $b$ are always labeled with $3$ and the light quarks are labeled with $1$ or $2$ as shown by Fig.~\ref{fig:w2to2} (a) and (b), i.e. the labels $i,j=1,2$ denote the light quarks in $su \to ud $, the relevant operator in a non-relativistic form can be written as:
\begin{equation}
\begin{aligned}
H_{W,2\to 2}^{PC,b}=&\kappa\frac{G_F}{\sqrt2}\frac{V_{us}V_{ud}}{(2\pi)^3} \hat\tau_1^{(+)} \hat\nu^{(-)}_2 \delta^3(\bm p'_1+\bm p'_2-\bm p_1-\bm p_2)\left( 1- \langle s_{z,1}'|\bm \sigma_1|s_{z,1}\rangle \langle s_{z,2}'| \bm \sigma_2 |s_{z,2}\rangle \right),
\\
%
H_{W,2\to 2}^{PV,b}=&\kappa\frac{G_F}{\sqrt 2} \frac{V_{us}V_{ud}}{(2\pi)^3}\hat\tau_1^{(+)} \hat\nu^{(-)}_2\delta^3(\bm p'_1+\bm p'_2-\bm p_1-\bm p_2) \\
&{}\times \left\{ -(\langle s_{z,1}'|\bm \sigma_1 |s_{z,1}\rangle - \langle s_{z,2}'|\bm\sigma_2 |s_{z,2}\rangle) \left [\left(\frac{\bm p_1}{2m_1}-\frac{\bm p_2}{2m_2}\right )+\left (\frac{\bm p'_1}{2m_1'}-\frac{\bm p'_2}{2m_2'}\right )\right ] \right.  \\
&{}+\left . i(\langle s_{z,1}'|\bm \sigma_1 |s_{z,1}\rangle \times \langle s_{z,2}'|\bm\sigma_2 |s_{z,2}\rangle) \left [\left (\frac{\bm p_1}{2m_1}-\frac{\bm p_2}{2m_2}\right )-\left (\frac{\bm p'_1}{2m_1'}-\frac{\bm p'_2}{2m_2'}\right)\right ] \right\},
\end{aligned}
\end{equation}
where $s_i \ (i=1,2)$ and $m_i$ are the spin and mass of the $i$-th quark, respectively; $\hat\tau$ and $\hat\nu$ are the flavor-changing operators, namely, $\hat\tau_1^{(+)}s_1=u_1,~\hat\nu_2^{(-)}u_2=d_2$. The symmetry factor $\kappa$ equals to $2$ because  $i$ could be $1$ or $2$ and $j$ will be determined when $i$ is assigned.

If the charm quark also participate in the interaction via $sc\to cd$, the flavor transition can be regarded as $s\to d$ in the approximation of contact interaction. Then, the complete operator can be written as:
\begin{equation}
\begin{aligned}
H_{W,2\to 2}^{PC,c}=&\kappa\frac{G_F}{\sqrt2}\frac{V_{cs}V_{cd}}{(2\pi)^3} \hat\alpha_1^{(-)} \delta^3(\bm p'_1+\bm p'_c-\bm p_1-\bm p_c)\left( 1- \langle s_{z,1}'|\bm \sigma|s_{z,1}\rangle \langle s_{z,c}'| \bm \sigma |s_{z,c}\rangle \right),
\\
%
H_{W,2\to 2}^{PV,c}=&\kappa\frac{G_F}{\sqrt 2} \frac{V_{cs}V_{cd}}{(2\pi)^3}\hat\alpha_1^{(-)} \delta^3(\bm p'_1+\bm p'_c-\bm p_1-\bm p_c) \\
&{}\times \left\{ -(\langle s_{z,1}'|\bm \sigma |s_{z,1}\rangle - \langle s_{z,c}'|\bm\sigma |s_{z,c}\rangle) \left [\left(\frac{\bm p_1}{2m_1}-\frac{\bm p_c}{2m_c}\right )+\left (\frac{\bm p'_1}{2m_1'}-\frac{\bm p'_c}{2m_c'}\right )\right ] \right.  \\
&{}+\left . i(\langle s_{z,1}'|\bm \sigma |s_{z,1}\rangle \times \langle s_{z,c}'|\bm\sigma |s_{z,c}\rangle) \left [\left (\frac{\bm p_1}{2m_1}-\frac{\bm p_c}{2m_c}\right )-\left (\frac{\bm p'_1}{2m_1'}-\frac{\bm p'_c}{2m_c'}\right)\right ] \right\},
\end{aligned}
\end{equation}
where $\hat\alpha_i$ is the flavor operator and $\hat\alpha_1^{(-)}s_1=d_1$; $\kappa$ is the symmetry factor which equals to 2.

\subsubsection{Quark-meson couplings in the chiral quark model}
The chiral quark model~\cite{Manohar:1983md} for the pion production is employed and the tree-level Hamiltonian is written as:
\begin{align}
H_m=\sum_j \int d \bm x \frac{1}{f_m}\bar q_j(\bm x) \gamma_\mu^j \gamma_5^j q_j(\bm x) \partial^\mu \phi_m(\bm x)   
\label{equ:Hpi},
\end{align}
where $f_m$ is the pseudoscalar meson decay constant; $q_j(\bm x)$ is the $j$-th quark field in the baryon and $\phi_m$ represents the meson field. In the chiral quark model the pseudoscalar meson only couples to the light quarks. With the fixed $j$, the non-relativistic form of the above equation can be expressed as:
\begin{align}
H_m=\frac{\kappa'}{\sqrt{(2\pi)^3 2\omega_m}} \frac{1}{f_m}\left[\omega_m\left(\frac{\bm\sigma\cdot \bm p^1_f}{2m_f}+\frac{\bm\sigma\cdot \bm p^1_i}{2m_i}\right) -\bm \sigma \cdot \bm k\right] \hat I^1_m \delta^3(\bm p^1_f+\bm k-\bm p_i^1),
\end{align}
where $\kappa'=2$ is the symmetry factor; $\omega_m$ and $\bm k$ are the energy and momentum of the pseudoscalar meson in the rest frame of the initial state, respectively; $\bm p^1_i$ and $\bm p^1_f$ are the initial and final momentum of the first quark, respectively; and $\hat I^1_m$ is the corresponding isospin operator for producing the pseudoscalar via its interaction with the first active quark within the baryon. For the production of the pion the isospin operator is written as:
\begin{align}
\hat I^1_\pi=\begin{dcases}
b^\dag_u b_d   &\text{for}~ \pi^-,\\
b^\dag_d b_u   &\text{for}~ \pi^+,\\
\frac{1}{\sqrt2}\left[b^\dag_u b_u- b^\dag_d b_d \right]    &\text{for}~ \pi^0,
\end{dcases}
\end{align}
where $b^\dag_{u/d}$ and $b_{u/d}$ are the creation and annihilation operators for the $u$ and $d$ quarks.

\subsection{Amplitudes and decay width}
With the operators and wave functions constructed in the quark model (see the Appendix) we formulate the transition amplitudes for the HQCWD processes in this subsection.  For convenience, the initial baryon and final baryon are noted as $B_i(\bm P_i;J_i,J^z_i)$ and $B_f(\bm P_f;J_f,J^z_f)$, respectively. The intermediate states in the pole terms are labeled as $B_m(\bm P_m;J_m,J^z_m)$ with $\bm P_m$ and $J^z_m$ determined by the conservation of three-vector momentum and angular momentum at the weak interaction vertex. The pion is labeled as $M_\pi(\bm k)$. The calculation is performed in the rest frame of the initial baryon, i.e. $\bm P_f=-\bm k$.

The matrix elements $\mathcal M$ collect contributions from different processes. For the pole terms, taking the A-type process as an example, the amplitude can be written as:
\begin{equation}
\mathcal M_\text{Pole,A}^{J_f,J_f^z;J_i,J_i^z}=\mathcal M_{\text{Pole,A};PC}^{J_f,J_f^z;J_i,J_i^z}+
\mathcal M_{\text{Pole,A};PV}^{J_f,J_f^z;J_i,J_i^z},
\end{equation}
where
\begin{equation}
\begin{aligned}\label{equ:amp}
&\begin{multlined}
\mathcal M_{\text{Pole,A};PC}^{J_f,J_f^z;J_i,J_i^z} \\
=\la B_f(\bm P_f;J_f,J_f^z) \middle| H_\pi \middle| B_m(\bm P_i;J_i,J_i^z) \ra \frac{i}{ \slashed p_{B_m}-m_{B_m} +i \frac{\Gamma_{B_m}}{2} } \la B_m(\bm P_i;J_i,J_i^z) \middle| H^{PC}_{W,2\to2} \middle| B_i(\bm P_i;J_i,J_i^z) \ra
\end{multlined}
\\
&\begin{multlined}
\mathcal M_{\text{Pole,A};PV}^{J_f,J_f^z;J_i,J_i^z} \\
= \la B_f(\bm P_f;J_f,J_f^z) \middle| H_\pi  \middle| B'_m(\bm P_i;J_i,J_i^z) \ra  \frac{i}{ \slashed p_{B'_m}-m_{B'_m} +i \frac{\Gamma_{B'_m}}{2} } \la B'_m(\bm P_i;J_i,J_i^z) \middle|  H^{PV}_{W,2\to2} \middle| B_i(\bm P_i;J_i,J_i^z) \ra,
\end{multlined}
\end{aligned}
\end{equation}
where $\left. \middle|B_m(\bm P_i;J_i,J_i^z)\ra$ and $\left. \middle|B'_m(\bm P_i;J_i,J_i^z)\ra$ denote the intermediate baryon states of $J^P=1/2^+$ and $1/2^-$, respectively, and $H_\pi$ means $\hat I^j_\pi$ is taken for $H_m$. In the calculation, we take the non-relativistic form for the propagators~\cite{Niu:2020gjw,Richard:2016hac}:
\begin{align}
\label{eq:propagator}
\frac{1}{\slashed p-m+ i\Gamma/2}\cong \frac{2m }{p^2-m^2+i \Gamma m}.
\end{align}
$\Gamma$ is the width of the intermediate baryon states.

The DPE and CS amplitudes have the same forms, respectively:
\begin{align}
\mathcal M_{CS}^{J_f,J_f^z;J_i,J_i^z}&=\langle B_f(\bm P_f;J_f,J_f^z);M(\bm k)|H_{W,1\to3}|B_c(\bm P_i;J_i,J_i^z) \rangle,\\
\mathcal M_{DPE}^{J_f,J_f^z;J_i,J_i^z}&=\langle B_f(\bm P_f;J_f,J_f^z);M_\pi(\bm k)|H_{W,1\to3}|B_c(\bm P_i;J_i,J_i^z) \rangle.
\end{align}
The main difference between these two processes is that the rearrangement of the initial quarks. This leads to the difference of the momentum conservation for the convolution of the spatial wave functions and a color-suppression factor. The calculation of the flavor and spin part can be found in our previous work~\cite{Niu:2020gjw}.

Using the normalization convention for the wave functions given in Appendix~\ref{app:wavefunction}, the decay width is expressed as:
\begin{align}
\Gamma(A\to B+C)=8\pi^2\frac{|\bm k|E_B E_C}{M_A}\frac{1}{2 J_A+1} \sum_\text{spin}|\mathcal M|^2,
\end{align}
where $J_A$ is the spin of the initial state and $\mathcal M$ is defined by
\begin{align}
\delta^3(\bm P_A -\bm P_B-\bm P_C)\mathcal M\equiv \langle BC | H_I|A \rangle.
\end{align}

\section{Numerical Results and discussions}\label{results}
\subsection{Parameters and inputs}

\begin{table}[htbp]
\centering
\caption{The intermediate states considered for the PC processes of pole terms.}
\begin{ruledtabular}
\begin{tabular}{lcccc}
Processes &$\Xi_c^+\to \Lambda_c \pi^0$ &$\Xi_c^0\to \Lambda_c \pi^-$ &$\Xi_b^0\to \Lambda_b \pi^0$&$\Xi_b^-\to \Lambda_b \pi^-$  \\
\hline
A-type &$\Sigma_c^+$,$\Lambda_c^+$ &$\Sigma_c^0$           &$\Sigma_b^0$, $\Lambda_b^0$ & $\cdots$ \\
B-type &$\Xi_c^+$, $\Xi'^+_c$      & $\Xi_c^+$, $\Xi'^+_c$ &$\Xi_b^0$, $\Xi'^0_b$       &$\Xi_b^0$, $\Xi'^0_b$\\
\end{tabular}
\end{ruledtabular}
\label{tab:states}
\end{table}

\begin{table}[htbp]
  \centering
  \caption{The masses of the antitriplet states (in unit of GeV). Only the central values of the masses are listed.} 
  \begin{ruledtabular}
    \begin{tabular}{c|llll|llll|llll|llll}
\multicolumn{1}{c|}{\multirow{2}[0]{*}{States}} & \multicolumn{4}{c|}{$\Lambda_c$} & \multicolumn{4}{c|}{$\Xi_c^+$}  & \multicolumn{4}{c|}{$\Lambda_b$} & \multicolumn{4}{c}{$\Xi_b^0/\Xi_b^-$}  \\
&$|^2 S\rangle$ &$|^2 P_\lambda\rangle$ &$|^2 P_\rho\rangle$ &$|^4 P_\rho\rangle$ 
&$|^2 S\rangle$ &$|^2 P_\lambda\rangle$ &$|^2 P_\rho\rangle$ &$|^4 P_\rho\rangle$
&$|^2 S\rangle$ &$|^2 P_\lambda\rangle$ &$|^2 P_\rho\rangle$ &$|^4 P_\rho\rangle$
&$|^2 S\rangle$ &$|^2 P_\lambda\rangle$ &$|^2 P_\rho\rangle$ &$|^4 P_\rho\rangle$ \\
\hline
PDG~\cite{Zyla:2020zbs} 
& 2.286 & 2.592  &$\cdots$&$\cdots$  
& 2.468 & 2.792  &$\cdots$&$\cdots$
& 5.62  & 5.912  &$\cdots$&$\cdots$
& 5.792 &$\cdots$&$\cdots$&$\cdots$ \\ 
Ref.~\cite{Copley:1979wj}
& 2.26  & 2.51  &$\cdots$&$\cdots$
&$\cdots$&$\cdots$&$\cdots$&$\cdots$ 
&$\cdots$&$\cdots$&$\cdots$&$\cdots$
&$\cdots$&$\cdots$&$\cdots$&$\cdots$  \\
Ref.~\cite{Yoshida:2015tia} 
& 2.285 & 2.628 & 2.890 & 2.933  
&$\cdots$&$\cdots$&$\cdots$&$\cdots$ 
& 5.618 & 5.938 & 6.236 & 6.273  
&$\cdots$&$\cdots$&$\cdots$&$\cdots$   \\
Ref.~\cite{Ebert:2011kk} 
& 2.286 & 2598  &$\cdots$&$\cdots$
& 2.476 & 2.792 &$\cdots$&$\cdots$
& 5.62  & 5.93  &$\cdots$&$\cdots$
& 5.803 & 6.12  &$\cdots$&$\cdots$ \\
Ref.~\cite{Chen:2016iyi} 
& 2.286 & 2.614 &$\cdots$&$\cdots$
& 2.47  & 2.793 &$\cdots$&$\cdots$
&$\cdots$&$\cdots$&$\cdots$&$\cdots$
&$\cdots$&$\cdots$&$\cdots$&$\cdots$ \\
Ref.~\cite{Roberts:2007ni}\footnotemark[1]
&$\cdots$&$\cdots$&$\cdots$&$\cdots$
&2.492  &2.763  &$\cdots$&$\cdots$
&$\cdots$&$\cdots$&$\cdots$&$\cdots$ 
&5.844  &6.108  &$\cdots$&$\cdots$\\
Used
&2.286  &2.592  &2.890  &2.933              
&2.468  &2.792  &2.990  &3.030              
&5.62   &5.912  &6.236  &6.273            
&5.792  &6.100  &6.230\cite{Aaij:2020fxj}  &6.270     \\
    \end{tabular}%
   \footnotetext[1]{~Only the unmixed data are used.}
    \end{ruledtabular}
\label{tab:antitriplet_mass}
\end{table}%

\begin{table}[htbp]
  \centering
  \caption{The masses of the sextet states (in unit of GeV). Only the central values of the masses are listed.}
\begin{ruledtabular}
    \begin{tabular}{c|llll|llll|llll|llll}
    \multicolumn{1}{c|}{\multirow{2}[0]{*}{States}} & \multicolumn{4}{c|}{$\Sigma_c^0/\Sigma_c^+$} & \multicolumn{4}{c|}{$\Xi'^+_c$} & \multicolumn{4}{c|}{$\Sigma_b^0$} & \multicolumn{4}{c}{$\Xi'^0_b$} \\
&$|^2 S\rangle$ &$|^2 P_\lambda\rangle$ &$|^2 P_\rho\rangle$ &$|^4 P_\lambda\rangle$
&$|^2 S\rangle$ &$|^2 P_\lambda\rangle$ &$|^2 P_\rho\rangle$ &$|^4 P_\lambda\rangle$ 
&$|^2 S\rangle$ &$|^2 P_\lambda\rangle$ &$|^2 P_\rho\rangle$ &$|^4 P_\lambda\rangle$ 
&$|^2 S\rangle$ &$|^2 P_\lambda\rangle$ &$|^2 P_\rho\rangle$ &$|^4 P_\lambda\rangle$ \\
\hline
PDG~\cite{Zyla:2020zbs} 
& 2.454 &$\cdots$&$\cdots$&$\cdots$  
& 2.578 &$\cdots$&$\cdots$&$\cdots$
& 5.816 &$\cdots$&$\cdots$&$\cdots$
& 5.935 &$\cdots$&$\cdots$&$\cdots$ \\
Ref.~\cite{Copley:1979wj} 
& 2.44  & 2.76\footnotemark[1]  &$\cdots$&$\cdots$
&$\cdots$&$\cdots$&$\cdots$&$\cdots$
&$\cdots$&$\cdots$&$\cdots$&$\cdots$
&$\cdots$&$\cdots$&$\cdots$&$\cdots$ \\
Ref.~\cite{Yoshida:2015tia} 
& 2.460 & 2.802 & 2.909 & 2.826   
&$\cdots$&$\cdots$&$\cdots$&$\cdots$ 
& 5.823 & 6.127 & 6.246 & 6.135  
&$\cdots$&$\cdots$&$\cdots$&$\cdots$  \\
Ref.~\cite{Ebert:2011kk} 
& 2.443 & 2.713 &$\cdots$& 2.799     
& 2.579 & 2.854 &$\cdots$& 2.936 
& 5.808 & 6.095 &$\cdots$& 6.101 
& 5.936 & 6.227 &$\cdots$& 6.233 \\
Ref.~\cite{Chen:2016iyi} 
& 2.456 & 2.702 &$\cdots$& 2.765     
& 2.579 & 2.839 &$\cdots$& 2.900   
&$\cdots$&$\cdots$&$\cdots$&$\cdots$       
&$\cdots$&$\cdots$&$\cdots$&$\cdots$  \\
Ref.~\cite{Roberts:2007ni} \footnotemark[2]
&$\cdots$&$\cdots$&$\cdots$&$\cdots$      
&2.529  &2.859  &$\cdots$&$\cdots$
&$\cdots$&$\cdots$&$\cdots$&$\cdots$       
&5.985  &6.192  &$\cdots$&$\cdots$ \\
Used 
&2.454  &2.802  &2.909  &2.826     
&2.578  &2.880  &2.980  &2.940     
&5.816  &6.097\cite{Aaij:2018tnn}  &6.246  &6.135      
&5.935  &6.220  &6.350  &6.250 \\
    \end{tabular}%
    \footnotetext[1]{~This is the mass for the composition of $-0.58|^4P_\lambda\rangle+0.79|^2P_\lambda\rangle$.}
    \footnotetext[2]{~Only the unmixed data are used.}
\end{ruledtabular}
  \label{tab:sextet_mass}%
\end{table}%

\begin{table}[htbp]
\centering
\caption{The quark mass used in different references.}
\begin{ruledtabular}
\begin{tabular}{cccccc}
inputs                     &$m_q$ (GeV) &$m_s$ (GeV) &$m_c$ (GeV) &$m_b$ (GeV)  \\
\hline
Ref.~\cite{Yoshida:2015tia}&$0.30$        &$0.51$        &$1.75$        &$5.11$      \\
Ref.~\cite{Roberts:2007ni} &$0.28$        &$0.56$        &$1.82$        &$5.20$      \\
Ref.~\cite{Ebert:2011kk}   &$0.33$        &$0.50$        &$1.55$        &$4.88$      \\
Used                       &$0.30\pm0.06$ &$0.50\pm0.10$ &$1.80\pm0.36$ &$5.0\pm1.0$      \\
\end{tabular}\end{ruledtabular}
\label{tab:parameters}
\end{table}

In the NRCQM framework, the parameters and inputs are divided into two categories. The first one is the quark model parameters which includes the constituent quark masses and the spring constant $K$ for the quark potential. 
The other one contains the masses of the intermediate states in the pole terms. 
As discussed above, only the near-by ground states with 
quantum number $1/2^+$ and the first orbital states with quantum number $1/2^-$ are considered. 
In this work, similar to the treatment of the hadronic weak decay processes~\cite{Niu:2020gjw}, only the ground states with quantum number $1/2^+$ (collected in Tab.~\ref{tab:states}) and the first orbital states with quantum number $1/2^-$ (collected in Tabs.~\ref{tab:antitriplet_mass} and \ref{tab:sextet_mass}) are considered. 
The masses of the ground states of these single heavy baryons are well determined and can be found in PDG~\cite{Zyla:2020zbs}. In contrast, their excited states are far from well established. This will inevitably introduce uncertainties into the theoretical estimate of the partial decay widths. This makes the quark model approach useful since at least a systematic treatment can be applied and its validity can be tested by confronting the available experimental data. Predictions can also be made for those processes which can be connected to each other by either microscopic mechanisms or symmetries. 
In Tab.~\ref{tab:parameters} the constituent quark masses from the literature~\cite{Zyla:2020zbs,Yoshida:2015tia,Copley:1979wj,Ebert:2011kk,Chen:2016iyi,Roberts:2007ni} are listed. We adopt the common values with uncertainties of 20~\% for the quark masses. We will also examine later the sensitivity of the results to the quark model parameters.

The value of $K$ ranges from $0.02$  to $0.038$ $\text{GeV}^3$ as shown by Ref.~\cite{Nagahiro:2016nsx}. In this work, we do not intend to calculate the branching ratios with high accuracy. So we adopt $K=0.015$ $\text{GeV}^3$ for the charmed baryons to best describe the available data from LHCb~\cite{Aaij:2015yoy}, and $K=0.035$ $\text{GeV}^3$ for the bottomed baryons as a prediction. Also, we will examine the errors caused by the uncertainty later.

As mentioned earlier the spectrum for excited heavy baryons is not well established in experiment, although tremendous theoretical studies can be found in the literature~\cite{Copley:1979wj,Yoshida:2015tia,Nagahiro:2016nsx,Roberts:2007ni,Thakkar:2016dna,Shah:2016nxi,Shah:2016mig,Chen:2016iyi,Ebert:2011kk,Garcilazo:2007eh,Ebert:2007nw,Giacchini:2018gxp,SilvestreBrac:1996bg,Xiao:2020gjo,Faustov:2020gun,Zhang:2008pm,Anisimov:2008dz,Narodetskii:2008pn,Liu:2007fg}. Recently, the LHCb collaboration has measured several heavy flavor baryon resonances. For instance,  $m[\Xi_c(2923)^0],m[\Xi_c(2939)^0],m[\Xi_c(2965)^0]$\cite{Aaij:2020yyt},$m[\Xi_b(6227)^0]$\cite{Aaij:2020fxj} and $m[\Sigma_b(6097)]$\cite{Aaij:2018tnn}.
Although their quantum numbers have not been determined and there might be other possible interpretations, some of them could be natural candidates for the $P$-wave excitations.

By assigning some of these recently observed states as the first orbital excitations one can estimate the partner states based on the quark model prescription. For the low-lying states the harmonic oscillator potential is a reasonable approximation as shown by various works in the literature. Then, the energy of a state can be expressed as 
\begin{align} 
 E_N =(2n_\rho+l_\rho+3/2)\omega_\rho+(2n_\lambda+l_\lambda+3/2)\omega_\lambda \ .
\end{align}
For the first excited states $n_\rho=n_\lambda=0$ and $l_\rho$ or $l_\lambda=1$. It can be verified that the the energy splitting between ground states and first orbital excitations is about $300\sim 400$ MeV and the energy splitting between the $\rho$ and $\lambda$ type excited states is about $200\sim 300$ MeV. In contrast, the mass splitting between $|^2P_\rho\rangle$ and $|^2P_\lambda\rangle$ configurations comes from the spin-dependent part of the potential and is about $30\sim 40$ MeV. Thus, we can estimate the masses of those first orbital states as a multiplet of the spin parity.

In Table~\ref{tab:states} we list the $P$-wave intermediate states needed in these channels. In Tables~\ref{tab:antitriplet_mass} and \ref{tab:sextet_mass} we list their masses either by assigning some of the states from LHCb as the first $P$-wave orbital excitation states of the flavor anti-triplet or sextet, or by adopting some theoretical estimates.  Note that the widths of these states are usually small, especially for the $b$-baryon sector, and most of them are not measured. As an estimate, we take a universal width of 5 MeV for all the first orbital excitation states in our calculations.

It should be noted that this treatment of the intermediate states will also introduce uncertainties into the final results. However, one will see later (see Tabs.~\ref{tab:spin-PC-CS}, \ref{tab:spin-PC-DPE} and \ref{tab:amp-cal}) that a large number of states would be forbidden from contribution due to the heavy quark symmetry. As a result, the uncertainties introduced by the undetermined properties will be limited. Here the mass difference of particles in the same isospin multiplet has been neglected.

\subsection{Numerical results and discussions}
Proceeding to the numerical calculations we skip the details of the transition matrix elements which are presented in  Appendix~\ref{app:amplitudes}, but focus on several characters arising from this kind of transitions. First, it shows that the HQS can be well understood in the NRCQM framework. For the CS and DPE processes, the spin operators of the weak interactions guarantee that the spin matrix elements of the PC processes vanish. Note that spin wave function of the $\bar 3$ states is $\chi^\rho$ which means that the spin wave function of the light quarks is antisymmetric with spin-parity $0^+$. The average values of the spin operators can be extracted from Eq.~(\ref{eq:HW13}) which are listed in Tabs.~\ref{tab:spin-PC-CS} and \ref{tab:spin-PC-DPE}. It shows clearly that the spin transition matrix elements vanishes if the wave functions of the initial and final baryons are both $\rho$ type. 
\begin{table*}[ht]
  \renewcommand{\arraystretch}{1.3}
  \caption{The spin matrix elements for the parity-conserving transitions in the CS process. Note that the spin wave function of pion is omitted.}
 \begin{ruledtabular}\begin{tabular}{ccccc}
$\mathcal{O}^\text{spin}$
&$\langle\chi^\lambda_{\half,-\half}|\mathcal{O}^\text{spin}|\chi^\lambda_{\half,-\half}\rangle$
&$\langle\chi^\lambda_{\half,-\half}|\mathcal{O}^\text{spin}|\chi^\rho_{\half,-\half}\rangle$
&$\langle\chi^\rho_{\half,-\half}|\mathcal{O}^\text{spin}|\chi^\lambda_{\half,-\half}\rangle$
&$\langle\chi^\rho_{\half,-\half}|\mathcal{O}^\text{spin}|\chi^\rho_{\half,-\half}\rangle$ \\
\hline
$\langle s'_1|I|s_1 \rangle \langle s_5 \bar s_4|\sigma_z|0\rangle$
&$\dfrac{\sqrt2}{3}$ &$-\dfrac{1}{\sqrt{6}}$ &$-\dfrac{1}{\sqrt{6}}$ &$0$ \\
$\langle s'_1|\sigma_z|s_1 \rangle \langle s_5 \bar s_4|I|0\rangle$
&$\dfrac{\sqrt2}{3}$ &$\dfrac{1}{\sqrt{6}}$ &$\dfrac{1}{\sqrt{6}}$ &$0$ \\
$(\langle s'_1|\bm \sigma|s_1 \rangle \times\langle s_5 \bar s_4|\bm \sigma |0\rangle)_z$  &$0$ &$-\dfrac{2i}{\sqrt 6}$ &$\dfrac{2i}{\sqrt 6}$ &$0$ \\
    \end{tabular}\end{ruledtabular}
\label{tab:spin-PC-CS}
\end{table*}
\begin{table*}[ht]
  \renewcommand{\arraystretch}{1.3}
  \caption{The spin matrix elements for the parity-conserving transitions in the CS process. Note that the spin wave function of pion is omitted.}
 \begin{ruledtabular}\begin{tabular}{ccccc}
$\mathcal{O}^\text{spin}$
&$\langle\chi^\lambda_{\half,-\half}|\mathcal{O}^\text{spin}|\chi^\lambda_{\half,-\half}\rangle$
&$\langle\chi^\lambda_{\half,-\half}|\mathcal{O}^\text{spin}|\chi^\rho_{\half,-\half}\rangle$
&$\langle\chi^\rho_{\half,-\half}|\mathcal{O}^\text{spin}|\chi^\lambda_{\half,-\half}\rangle$
&$\langle\chi^\rho_{\half,-\half}|\mathcal{O}^\text{spin}|\chi^\rho_{\half,-\half}\rangle$ \\
\hline
$\langle s'_1|\sigma_z|s_1 \rangle \langle s_5 \bar s_4|I|0\rangle$
&$-\dfrac{2\sqrt2}{3}$ &$\dfrac{2}{\sqrt{6}}$ &$\dfrac{2}{\sqrt{6}}$ &$0$ \\
    \end{tabular}\end{ruledtabular}
\label{tab:spin-PC-DPE}
\end{table*}
\begin{table}[htbp]
\centering
\begin{ruledtabular}
\caption{Calculated results for exclusive terms and total amplitudes in unit of $10^{-9}~\text{GeV}^{-1/2}$. The pole terms for the $\Xi_c$ decays include both transitions where $c$ is either a spectator or not. It should be noted the tag is used to label the process which $c$ is not spectator for $\Xi_c$ decay processes.}\label{tab:amp-cal}
\begin{tabular}{cccrcrcrcr}
          &       & \multicolumn{2}{c}{$\Xi^+_c \to \Lambda_c \pi^0$} & \multicolumn{2}{c}{$\Xi_c^0 \to \Lambda_c \pi^-$} & \multicolumn{2}{c}{$\Xi_b^0 \to \Lambda_b \pi^0$} & \multicolumn{2}{c}{$\Xi_b^- \to \Lambda_b \pi^-$} \\
\hline \multirow{7}[0]{*}{PC} 
&Pole-A &$\Sigma_c^+$  &$103.95-15.85i$ &$\Sigma_c^0$  &$130.66-6.95i$ & $\Sigma_b^0$ &Spin(weak)&           &  \\
&       &$\Lambda_c^+$ &Isospin         &              &               & $\Lambda_b$  &Spin(CQM) &           &  \\          
&Pole-B & $\Xi_c^+$    &Spin(CQM)       & $\Xi_c^+$    &Spin(CQM)      & $\Xi_b^0$    &Spin(CQM) &$\Xi_b^0$  &Spin(CQM)  \\
&       & $\Xi'^+_c$   &$-2.90$         & $\Xi'^+_c$   &$-4.08$        & $\Xi'^0_b$   &Spin(weak)&$\Xi'^0_b$ &Spin(weak)  \\
& DPE   &              &Spin(weak)      &              &Spin(weak)     &              &Spin(weak)&           &Spin(weak)  \\
& CS    &              &Spin(weak)      &              &Spin(weak)     &              &Spin(weak)&           &Spin(weak)  \\
& Total &              &$101.05-15.85i$ &              &$126.59-6.95i$ &              &$0$       &           &$0$  \\
\hline   
\hline
    \multirow{15}[0]{*}{PV} 
&Pole-A &$\Sigma_c^+|^2 P_\rho\rangle$    &Spin(CQM) & $\Sigma_c^0|^2 P_\rho\rangle$    &Spin(CQM)& $\Sigma_b^0|^2 P_\rho\rangle$    &Spin(weak) &  &  \\
&       &$\Sigma_c^+|^2 P_\lambda\rangle$ &$1.79+0.014i$&$\Sigma_c^0|^2 P_\lambda\rangle$ &$2.56+0.021i$   & $\Sigma_b^0|^2 P_\lambda\rangle$ &Spatial    &  &  \\
&       &$\Sigma_c^+|^4 P_\rho\rangle$    &$-1.06$   & $\Sigma_c^0|^4 P_\rho\rangle$    &$-1.52$  & $\Sigma_b^0|^4 P_\rho\rangle$    &Spatial    &  &  \\
&       &$\Lambda_c^+|^2 P_\rho\rangle$   &Isospin   &                                  &         & $\Lambda_b|^2 P_\rho\rangle$     &Isospin    &  & \\
&       &$\Lambda_c^+|^2 P_\lambda\rangle$&Isospin   &                                  &         & $\Lambda_b|^2 P_\lambda\rangle$  &Isospin    &  & \\
&       &$\Lambda_c^+|^4 P_\rho\rangle$   &Isospin   &                                  &         & $\Lambda_b|^4 P_\rho\rangle$     &Isospin    &  & \\        
&Pole-B &$\Xi_c^+|^2 P_\rho\rangle$       &$-3.45-0.017i$&$\Xi_c^+|^2 P_\rho\rangle$    &$-4.89-2.38i$  & $\Xi_b^0|^2 P_\rho\rangle$       &$-8.27-0.036i$& $\Xi_b^0|^2 P_\rho\rangle$     &$-11.71-0.051i$  \\
&       &$\Xi_c^+|^2 P_\lambda\rangle$    &Spin(CQM) &$\Xi_c^+|^2 P_\lambda\rangle$     &Spin(CQM)& $\Xi_b^0|^2 P_\lambda\rangle$    &Spin(CQM)  & $\Xi_b^0|^2 P_\lambda\rangle$  &Spin(CQM)  \\
&       &$\Xi_c^+|^4 P_\rho\rangle$       &$-1.17$   &$\Xi_c^+|^4 P_\rho\rangle$        &$-1.65-0.013i$  & $\Xi_b^0|^4 P_\rho\rangle$       &$-3.71-0.014i$ & $\Xi_b^0|^4 P_\rho\rangle$     &$-5.26-0.20$  \\
&       &$\Xi'^+_c|^2 P_\rho\rangle$      &Spin(CQM) &$\Xi'^+_c|^2 P_\rho\rangle$       &Spin(CQM)& $\Xi'^0_b|^2 P_\rho\rangle$      &Spin(CQM)  & $\Xi'^0_b|^2 P_\rho\rangle$    &Spin(CQM)  \\
&       &$\Xi'^+_c|^2 P_\lambda\rangle$   &$0.43$    &$\Xi'^+_c|^2 P_\lambda\rangle$    &$0.60$   & $\Xi'^0_b|^2 P_\lambda\rangle$   &Spatial    & $\Xi'^0_b|^2 P_\lambda\rangle$ &Spatial  \\
&       &$\Xi'^+_c|^4 P_\lambda\rangle$   &$-0.31$   &$\Xi'^+_c|^4 P_\lambda\rangle$    &$-0.44$  & $\Xi'^0_b|^4 P_\lambda\rangle$   &Spatial    & $\Xi'^0_b|^4 P_\lambda\rangle$ &Spatial  \\
& DPE   &                                 &$0$       &                                  &$-9.73$ &                                  &$0$        &                                &$-9.80$  \\
& CS    &                                 &$3.40$    &                                  &$4.81$   &                                  &$4.54$     &                                &$6.42$    \\
& Total &                                 &$-0.38+0.014i$&                              &$-10.25-0.019i$ &                           &$-7.44-0.050i$ &                            &$-20.03-0.071i$  \\
\end{tabular}
\end{ruledtabular}
\end{table}

Concerning the pole terms, the transition amplitudes combine the weak and strong interaction vertices. Note that the strong interactions only involve the light quarks in these processes considered in this work, while the weak interactions can involve the heavy one in the charm sector. It means that the PC processes where the heavy quark behaves as a spectator are forbidden. This is actually the consequence of the HQS. When the heavy quark, i.e. $c$ or $b$, behaves as a spectator, the amplitudes will be determined by the dynamics of the light quarks via $0^+\to 0^+ 0^-$. Apparently, this transition is forbidden for the PC operators~\cite{Li:2014ada}. This selection rule is stringent for the $\Xi_b$ decays since the internal conversion processes involving the $b$ quark are highly suppressed. In contrast, the $c$ quark in the $\Xi_c$ decays can participate the weak interactions and allow the contributions from the PC operators. Another feature arising from the pole terms is that the amplitude given by the intermediate $\Sigma_c$ turns out to be dominant in $\Xi_c\to \Lambda_c\pi$.  This suggests an interesting prediction to be observed in experiment. Namely, the PC amplitudes via the intermediate $\Sigma_c$ is dominant in $\Xi_c\to \Lambda_c\pi$ while in $\Xi_b\to \Lambda_b\pi$ the amplitudes are dominated by the PV mechanism since the PC transitions are forbidden.

Combining the strong and weak couplings together and including the constraints from the symmetry selection rules, some of the intermediate states will be forbidden from contribution. Table~\ref{tab:amp-cal} collects all the amplitude informations where the labels ``Spin (weak)", ``Spin (CQM)", ``spatial" and ``Isospin" denote the transitions which are forbidden by the spin operators at the weak and strong interaction vertices, spatial operators and isospin operators, respectively.

All the calculated amplitudes are listed in Table~\ref{tab:amp-cal} and the detailed formulas can be found in Appendix~\ref{app:amplitudes}. It shows that for the decay of $\Xi_c\to\Lambda_c \pi$ the largest contributions come from the pole terms of $\Sigma_c$. This is because the mass of $\Sigma_c$ is very close to the mass of $\Xi_c$. Therefore, the amplitudes will be strongly enhanced by the propagator. Since the PC amplitudes vanish in $\Xi_b\to\Lambda_b\pi$ one anticipates from Table~\ref{tab:amp-cal} that the asymmetry parameters should be zero for the $\Xi_b$ decay processes. The asymmetry parameters for $\Xi_c$ decays should also be very small because the value of the PC amplitudes are one or two orders of magnitude larger than the PV ones.

\begin{table*}[ht]
  \renewcommand{\arraystretch}{1.3}
  \caption{The calculated branching ratios (last row) are compared with the available experimental data and other model results (in $\%$). Note that only the results for a constructive interference phase in Ref.~\cite{Gronau:2016xiq} are cited here, considering that $\mathcal{B}(\Xi_c^0\to \Lambda_c \pi^-)$ has a large value.}
 \begin{ruledtabular}\begin{tabular}{crrrr}
processes
&$\Xi_c^+\to \Lambda_c \pi^0$ &$\Xi_c^0\to \Lambda_c \pi^-$ &$\Xi_b^0\to \Lambda_b \pi^0$ &$\Xi_b^-\to \Lambda_b \pi^-$ \\
\hline
Exp. Data&$\cdots$&$0.55\pm0.20$~\cite{Aaij:2020wtg}&$\cdots$ &$0.19\pm0.07\sim0.57\pm0.21$~\cite{Aaij:2015yoy}
\\
MIT bag model\cite{Cheng:2015ckx}   &$0.0093$  &$0.0087$  &$0.059$       &$0.2$ \\
Diquark model\cite{Cheng:2015ckx}   &$\cdots$  &$\cdots$  &$0.25$        &$0.69$ \\
Duality\cite{Gronau:2015jgh}        &$\cdots$  &$\cdots$  &$\cdots$ &$0.63\pm 0.42$ \\
Current algebra\cite{Gronau:2016xiq}&$0.386\pm0.135$   &$0.194\pm0.07$   &$\cdots$      &$\cdots$ \\
Current algebra\cite{Li:2014ada}    &$\cdots$  &$\cdots$  &$1\sim4$      &$2\sim8$ \\
Current algebra\cite{Faller:2015oma}&$<0.6$    &$<0.3$    &$0.09-0.37$   &$0.19-0.76$\\
Our results                         &$1.11$    &$0.58$    &$0.017$       &$0.14$ \\
 \end{tabular}\end{ruledtabular}
\label{tab:br}
\end{table*}

The calculated branching ratios are listed in Table~\ref{tab:br} to be compared with the available experimental data and other model calculations. It shows that the calculated $\mathcal{B}(\Xi_c^0\to\Lambda_c\pi^-)=0.58\%$ is consistent with the experimental data, i.e. $(0.55\pm0.20)\%$\cite{Aaij:2020wtg}, within the experimental uncertainty. In our model, the decays of $\Xi_c\to \Lambda_c\pi$ are saturated by the A-type pole terms via the conversion subprocess $cs\to cd$ because the masses of $\Xi_c$ and the intermediate $\Sigma_c$ are very close. The $\Sigma_c$ propagator is almost on shell which leads to a strong enhancement of the transition amplitude. The difference between $\Xi_c^+\to\Lambda_c\pi^0$ and $\Xi_c^0\to\Lambda_c\pi^-$ is that the flavor wave function between $\pi^0$ and $\pi^-$, which results in a factor of $1/\sqrt{2}$ between the amplitudes for the $\pi^0$ and $\pi^-$ channel in Table~\ref{tab:amp-cal}. Taking into account that the life time of $\Xi_c^-$ is about four times of that of $\Xi_c^0$, we find that the branching ratio fraction $\mathcal{B}(\Xi_c^+\to\Lambda_c\pi^0)/\mathcal{B}(\Xi_c^0\to\Lambda_c\pi^-)\simeq 2$, 
which is consistent with the result of Ref.~\cite{Gronau:2016xiq}.
 Interestingly, we find that this ratio holds no matter the two transition processes $s\to u\bar u d$ and $cs\to cd$ interfere destructively or constructively. Note that Ref.~\cite{Gronau:2016xiq} also claimed that these two types of transitions have the same magnitude. It implies that the DPE process via $s\to u\bar u d$ should be much smaller than the conversion term via $cs\to cd$ plus the CS process via $s\to u\bar u d$, where the latter two processes can hold  $\Gamma(\Xi_c^+\to\Lambda_c\pi^0)/\Gamma(\Xi_c^0\to\Lambda_c\pi^-)\simeq 1/2$ due to the normalization of the flavor wavefunctions for $\pi^0$ and $\pi^-$. 

 A comparison of the $\Xi_c^+\to \Lambda_c \pi^0$, $\Xi_c^0\to \Lambda_c \pi^-$, $\Xi_b^0\to \Lambda_b \pi^0$, 
  and $\Xi_b^-\to \Lambda_b \pi^-$ branching ratios to other theoretical results, i.e. MIT bag model~\cite{Cheng:2015ckx}, diquark model~\cite{Cheng:2015ckx}, duality~\cite{Gronau:2015jgh}, 
  current algebra~\cite{Faller:2015oma, Gronau:2016xiq, Li:2014ada} and experimental data~\cite{Aaij:2020wtg, Aaij:2015yoy}
  are collected in Tab.~\ref{tab:br}. Since the results from the diquark model and duality are very limited, we focus on the comparison between ours and the results from MIT bag model~\cite{Cheng:2015ckx}
  and current algebra~\cite{Faller:2015oma, Gronau:2016xiq, Li:2014ada}. 
  
For the current algebra calculations~\cite{Faller:2015oma,Li:2014ada}, due to the pure $S$-wave transitions with $\Delta I=1/2$ one has $\mathcal{M}(\Xi_b^-\to\Lambda_b\pi^-)/\mathcal{M}(\Xi_b^0\to\Lambda_b\pi^0)=\sqrt 2$ which leads to the relation $\mathcal{B}(\Xi_b^-\to\Lambda_b\pi^-)/\mathcal{B}(\Xi_b^0\to\Lambda_b\pi^0)\approx 2$. This ratio arises from the flavor wave functions between the $\pi^0$ and $\pi^-$ channel.
In contrary, the ratio in our model is different. As mentioned earlier, in the $\Xi_b$ decays the $b$ quark always acts as a spectator. Meanwhile, the DPE contribution is the same order of magnitude as the pole terms and CS contributions.
It will introduce interferences which can change the partial width ratio of $1/2$ between the $\pi^0$ and $\pi^-$ channel (see Tab.~\ref{tab:amp-cal}).
      Taking into account the total width, we obtain $\mathcal{B}(\Xi_b^-\to\Lambda_b\pi^-)=(0.14\pm 0.073)\%$ which is consistent with the experimental measurement $(0.19\pm0.07)\sim(0.57\pm0.21)\%$~\cite{Aaij:2015yoy}. 
      The predicted branching ratio of $\Xi_b^0\to\Lambda_b\pi^0$ is $0.017\%$ which is an order of magnitude smaller than $\mathcal{B}(\Xi_b^-\to\Lambda_b\pi^-)$ due to the lack of contribution of DPE. This is of the same order as the results from Refs.~\cite{Cheng:2015ckx, Faller:2015oma}.

In the MIT bag model~\cite{Cheng:2015ckx}, the branching ratios of $\Xi_c$ are predicted to be about one order of magnitude smaller than that of $\Xi_b$, which is because of the short life time of $\Xi_c$ and the destructive contributions from the non-spectator $W$-exchange. Our model predicts differently as shown in Table~\ref{tab:br}, where the large branching ratios of $\Xi_c\to \Lambda_c\pi$ are due to the significant pole term contributions in the parity-conserving transitions. 


\begin{table*}[ht]
  \renewcommand{\arraystretch}{1.3}
  \caption{Uncertainties of the branching ratios (in \%) caused by the quark model parameters with 20\%
errors. }
 \begin{ruledtabular}\begin{tabular}{ccccc}
Input &$\Xi_c^+\to \Lambda_c \pi^0$ &$\Xi_c^0\to \Lambda_c \pi^-$ &$\Xi_b^0\to \Lambda_b \pi^0$ &$\Xi_b^-\to \Lambda_b \pi^-$ \\
\hline
Exp. Data&$\cdots$&$0.55\pm0.20$\cite{Aaij:2020wtg}&$\cdots$ &$0.19\pm0.07\sim0.57\pm0.21$\cite{Aaij:2015yoy}
\\
$m_q$   &$1.11\pm0.10$&$0.58\pm0.051$              &$0.017\pm0.0082$     &$0.14\pm0.033$ \\
$m_s$   &$1.11\pm0.17$ &$0.58\pm0.088$              &$0.017\pm0.0011$     &$0.14\pm0.0064$ \\
$m_c$   &$1.11\pm0.053$ &$0.58\pm0.027$             &$0.017\pm0$         &$0.14\pm0$ \\
$m_b$   &$1.11\pm0$    &$0.58\pm0$                 &$0.017\pm0$         &$0.14\pm0$ \\
$K$     &$1.11\pm0.34$ &$0.58\pm0.18$            &$0.017\pm0.012$      &$0.14\pm0.054$ \\
$R$     &$1.11\pm0$    &$0.58\pm0.002$                 &$0.017\pm0.0011$    &$0.14\pm0.036$ \\
Combined&$1.11\pm0.40$ &$0.58\pm0.21$              &$0.017\pm0.015$     &$0.14\pm0.073$ \\
    \end{tabular}\end{ruledtabular}
\label{tab:brerr}
\end{table*}

We also investigate the uncertainties by introducing $20\%$ errors to the quark model parameters and the results are listed in Table~\ref{tab:brerr}. It clearly shows that the branching ratios are sensitive to the light quark mass and the spring constant $K$, which actually determine the hadron wave functions.  It suggests that the hadronic weak decay is a sensitive probe for the hadron structures~\cite{Niu:2020gjw}.  
Another source of errors, which is not included in Table~\ref{tab:brerr}, is the masses of first orbital excitation states in the pole terms. At the moment, precise experimental data are still unavailable, while  theoretical estimates vary dramatically based on different methods. Moreover, we have not considered the state mixings due to lack of dynamical prescriptions. However, it should be stressed that the main features and outcomes from the quark model approach can still serve as a useful guidance for future experimental measurements.

Furthermore, we also estimate the uncertainties arising from the masses of the pole terms by the error transfer formulas. The uncertainty of a function $f(x)$ can be expressed in the first-order series approximation as 
\begin{align}
f(x)\pm \sqrt{\Delta x^2 f'(x)^2}.
\end{align}
Then, the error caused by the mass uncertainty $\Delta m$ of an intermediate state with mass $m$ can be estimated by 
\begin{align}
\label{eq:err}
\frac{2m}{s-m^2}\left(1\pm 
 \frac{\Delta m}{m} \frac{s+m^2}{s-m^2}\right),
\end{align}
where $s$ is the four-vector momentum squared which is known. It should be noted that since the widths of the intermediate states are usually narrow, the width effects are neglected here for simplicity.

Equation~(\ref{eq:err}) shows that the errors could be large if the intermediate states are close to the on-shell kinematic region. The masses of the ground states with $J^P=1/2^+$ are well determined by experiment. So we will not discuss the pole terms in the PC processes, but focus on those in the PV processes. As mentioned earlier, most of those $1/2^-$ singly heavy baryons have not been confirmed in experiment. For the $\Xi_c$ decay processes, the main contribution comes from the PC pole terms. Therefore, the uncertainties caused by the $1/2^-$ intermediate baryons are very limited. In this sense we find that our model calculation can give a reliable estimate of the partial decay widths for $\Xi_c\to \Lambda_c\pi$.  

Different from the $\Xi_c$ decays, the PV pole terms play an essential role in the $\Xi_b$ decays. With $\Delta m=100$ MeV as the input for the first orbital excitation states, the uncertainties can be determined by the mass ratio $\sqrt{s}/m=M(\Lambda_b)/m(\Sigma_b(\frac 12^-))=5.8/6.3\approx 0.90$ for the B-type pole terms. With $\left|\frac{\Delta m}{m} \frac{M^2+m^2}{M^2-m^2}\right|\approx 20\%$ one estimates $\Delta \mathcal{B}(\Xi_b^0 \to \Lambda_b \pi^0)\approx 60\%$ and $\Delta \mathcal{B}(\Xi_b^- \to \Lambda_b \pi^-)\approx 30\%$. It shows that the uncertainties of the intermediate states have more significant effects on $\Xi_b^0\to \Lambda_b\pi^0$ compared with $\Xi_b^- \to \Lambda_b \pi^-$. This is due to the sizeable contributions from the DPE process in $\Xi_b^- \to \Lambda_b \pi^-$ which can decrease the effects of the mass uncertainties of the intermediate states. A joint analysis of these two processes may provide additional constraints on the intermediate $1/2^-$ states.

\section{Summary}\label{summary}

In this work we present a systematic study of the heavy quark conserving weak decay processes, $\Xi_c^+\to\Lambda_c\pi^0$, $\Xi_c^0\to\Lambda_c\pi^-$, $\Xi_b^0\to\Lambda_b\pi^0$ and $\Xi_b^-\to\Lambda_b\pi^-$, in the framework of non-relativistic constituent quark model.  Our calculations show that these four processes cannot be trivially related to each other by either flavor symmetry or HQS. For the $\Xi_c$ and $\Xi_b$ decays, there are two main dynamic differences which can drastically change the pattern expected by the flavor symmetry or HQS. In $\Xi_c\to \Lambda_c\pi$ the PC pole terms are found dominant and the DPE contributions are rather small. It predicts sizeable branching ratios for both $\Xi_c^+\to\Lambda_c\pi^0$ and $\Xi_c^0\to\Lambda_c\pi^-$. Moreover, it shows that the charm quark can actually participate the weak decay but still keep the heavy quark conserved, i.e. via $cs\to dc$. This suggests that some HQS breaking effects are anticipated in the $\Xi_c$ decays. In contrast, the bottom quark in $\Xi_b$ decays is just a spectator, thus, the HQS will be better respected. As a consequence of the HQS, the PC transitions vanish in $\Xi_b\to \Lambda_b\pi$. Meanwhile, the non-factorizable pole terms and CS transitions are compatible with the DPE. This makes the relation between $\Xi_b^0\to\Lambda_b\pi^0$ and $\Xi_b^-\to\Lambda_b\pi^-$ more complicated than that based on the simple flavor symmetry. In fact, in the $\Xi_b$ decays, it is the PV component dominates the transition amplitudes. Interestingly, although $\Xi_c$ and $\Xi_b$ are so different in terms of the transition mechanisms, they both have small values for the asymmetry parameter due to the dominance of the PC and PV components respectively for the $\Xi_c$ and $\Xi_b$ decays. This can be investigated in experiment in the future. 

We also make an analysis of the uncertainties of our model calculations. We find that the major uncertainties are originated from the light constituent quark mass and spring constant $K$, and the mass of first orbital excitation states in the pole terms. This indicates that the weak decays are ideal probes for hadron internal structures. We find that with about $20\%$ uncertainties for the quark model parameters, we can still obtain reasonably accurate results for the branching ratios. This makes the predictions for $\Xi_c^+\to \Lambda_c\pi^0$ and $\Xi_b^0\to \Lambda_b\pi^0$ extremely interesting, and they can be search for in experiment, e.g. at LHCb.

\begin{acknowledgments}
Useful discussions with Qi-fang L\"u on the single heavy baryons mass spectrum are acknowledged. We thank Yu Lu for the help on solving a technical problem in the analytic computations. This work is supported, in part, by the National Natural Science Foundation of China (Grant Nos. 11425525 and 11521505),  DFG and NSFC funds to the Sino-German CRC 110 ``Symmetries and the Emergence of Structure in QCD'' (NSFC Grant No. 12070131001, DFG Project-ID 196253076), Strategic Priority Research Program of Chinese Academy of Sciences (Grant No. XDB34030302), National Key Basic Research Program of China under Contract No. 2020YFA0406300. Q.W. is also supported by Guangdong Major Project of Basic and Applied Basic Research No.~2020B0301030008, the NSFC under Grant No.~12035007, the Science and Technology Program of Guangzhou No.~2019050001, and Guangdong Provincial Funding under Grant No.~2019QN01X172.

\end{acknowledgments}

\begin{appendix}
\section{Coordinates and spatial wave functions}\label{App:wavefuncion}
\subsection{Jacobi coordinates}
\label{app:JAC}

We briefly introduce the Jacobi coordinates of $N$ particles. The motion of the $N$ particles can be separated into two types. The first one is the center-of-mass (c.m.) motion and the other one is the  relative motion between the $i+1$-th ($1\leq i \leq N-1$) particle and the c.m. motion of the first $i$ particles. Then, the Jacobi coordinates can be defined as:
\begin{align}
\label{eq:JCCr}
\bm R   &=\frac{\sum_{i=1}^N m_i \bm r_i}{M}, \\
\bm R_i &=\frac{M_i}{\sqrt{M_i^2+M_i^{(2)}}} \left[ \sum_{j=1}^i\frac{ m_j \bm r_j}{M_i}- \bm r_{i+1} \right ],~~~i=1,\cdots,N-1,
\end{align}
where 
\begin{align}
M=\sum_{i=1}^N m_i,~M_i= \sum_{j=1}^{i} m_j,~M_i^{(2)}= \sum_{j=1}^{i} m_j^2,~~~ i=1,\cdots,N-1.
\end{align}
$\bm r_i$ and $m_i$ denote the position and mass of the $i$-th particle. The factor $\frac{M_i}{\sqrt{M_i^2+M_i^{(2)}}}$ can be considered as the normalization factor of $\bm R_i$.
The reduce mass $\mu_i$ is defined by
\begin{align}
\label{eq:reducemass}
\sum_{i=1}^N m_i \bm r_i^2=M \bm R^2+ \sum_{i=1}^{N-1}\mu_i \bm R_i^2.
\end{align}
Then, we can easily obtain the expression of $\mu_i$
\begin{align}
\mu_i=\frac{m_{i+1}\left(M_i^2+M_i^{(2)}\right)}{M_i M_{i+1}},~~~ i=1,\cdots,N-1 .
\end{align}
The momentum which are the conjugate variables of $\bm R$ and $\bm r_i$ are defined as:
\begin{align}
\label{eq:JCCp}
\bm P=M \bm \dot{R},~\bm P_i =\mu_i \bm \dot{R_i},
\end{align}
where $\bm P$ is the total momentum of the $N$ particle system, i.e. the overall c.m. momentum. From Equation.~(\ref{eq:reducemass}), we can obtain
\begin{align}
\sum_{i=1}^N\frac{\bm p_i^2}{2 m_i^2}=\frac{\bm P^2}{2 M}+\sum_{i=1}^{N-1}\frac{\bm P_i^2}{2 \mu_i^2},
\end{align}
where $\bm p_i =m_i \bm \dot{r_i}$. This indicates that the total kinetic energy of the $N$ particle system is conserved as expected.

\subsection{Spatial wave functions for three-body system}
In the NRCQM the harmonic potential is adopted for the spin-independent part of the quark interactions, while the spin-dependent potential is minimized to be an anharmonic correction to the harmonic part. We construct the wave functions following the convention of Isgur {\it et al.}~\cite{Isgur:1978xj}. The general form of Hamiltonian for a three-quark baryon system is:
\begin{equation}
\label{eq:Ham}
H=\sum_{i=1}^3 \frac{\bm p_i^2}{2 m_i}+\frac{1}{2}K \sum_{i<j} (\bm r_i-\bm r_j)^2,
\end{equation}
where $K$ is the the spring constant. With Eqs.~(\ref{eq:JCCr}) and (\ref{eq:JCCp}), we can obtain
\begin{equation}
\label{eq:transition}
\left\{
\begin{aligned}
\bm R &= \frac{1}{M}\left( m_1 \bm r_1+m_2 \bm r_2+m_3 \bm r_3 \right) \\
\bm \rho & \equiv \bm R_1=\frac{1}{\sqrt 2}\left(\bm r_1 -\bm r_2 \right) \\
\bm \lambda &\equiv \bm R_2= \frac{m_1\bm r_1+m_2\bm r_2-(m_1+m_2)\bm r_3}{\sqrt{m_1^2+m_2^2+(m_1+m_2)^2}}
\end{aligned}\right. ,~
\left\{
\begin{aligned}
\bm P&= \bm p_1+\bm p_2+\bm p_3\\
\bm p_\rho&\equiv \bm P_1= \frac{\sqrt 2}{m_1+m_2} \left( m_2 \bm p_1- m_1 \bm p_2 \right)\\
\bm p_\lambda&\equiv \bm P_2=
\frac{2(m_1^2+m_2^2+m_1 m_2)\left[m_3\bm p_1+m_3\bm p_2-(m_1+m_2)\bm p_3 \right]}{M(m_1+m_2)\sqrt{m_1^2+m_2^2+(m_1+m_2)^2}} 
\end{aligned}\right. \ .
\end{equation}
Then, the Hamiltonian can be rewritten as:
\begin{align}
\label{eq:HamJ}
H=\frac{\bm P^2}{2M}+\frac{\bm p_\rho^2}{2 m_\rho}+\frac{\bm p_\lambda^2}{2 m_\lambda}+\frac{1}{2} m_\rho \omega_\rho^2 \bm \rho^2 +\frac{1}{2} m_\lambda \omega_\lambda^2 \bm \lambda^2,
\end{align}
where $M$ is the total mass. The reduced masses of the $\rho$ and $\lambda$ degrees of freedom are defined as:
\begin{eqnarray}
m_\rho &\equiv &\mu_1=\frac{2 m_1 m_2}{m_1+m_2}, \\
m_\lambda &\equiv &\mu_2=\frac{2m_3(m_1^2+m_2^2+m_1m_2)}{M(m_1+m_2)}.
\end{eqnarray}
The frequencies of the corresponding harmonic oscillators are expressed as:
\begin{equation}
\omega_\rho=\sqrt{3K/m_\rho}, \ \ \omega_\lambda=\sqrt{3K/m_\lambda}.
\end{equation}

Finally, the spatial wave functions of the harmonic oscillator basis can be easily obtained~\cite{LeYaouanc:1988fx,Pervin:2005ve,Zhong:2007gp}. In the coordinate space, the eigen spatial wave functions are:
\begin{align}
\Psi_{N,L,L_z}(\bm R,\bm \rho,\bm\lambda)
=\frac{1}{(2\pi)^{3/2}}
\exp\left(-i \bm P\cdot \bm R\right)
\sum_m \langle l_\rho,m;l_\lambda,L_z-m|L,L_z \rangle\psi^{\alpha_\rho}_{n_\rho l_\rho m }(\bm \rho)\psi^{\alpha_\lambda}_{n_\lambda l_\lambda L_z-m }(\bm \lambda),
\end{align}
where $N$ stands for $\{n_\rho,l_\rho;n_\lambda,l_\lambda\}$, and
\begin{align}
\psi^\alpha_{nlm}(\bm r)=\left[\frac{2\,n!}{(n+l+1/2)!} \right]^{1/2}\alpha^{l+3/2}
\exp\left(-\frac{\alpha^2\bm r^2}{2}\right)L_n^{l+1/2}(\alpha^2\bm r^2)
\mathcal{Y}_{lm}(\bm r),
\end{align}
where $\bm P$ is the total momentum of the three quark system. The function $L_n^\nu(x)$ is the generalized Laguerre polynomial, and $\alpha_\rho$ and $\alpha_\lambda$ are the harmonic oscillator strengths defined by
\begin{align}
\label{eq:hos}
\alpha_\rho^2=m_\rho \omega_\rho, \ \ \alpha_\lambda^2=m_\lambda \omega_\lambda .
\end{align}
In the momentum space, the wave functions are:
\begin{align}
\Psi_{N L L_z}(\bm P,\bm p_\rho,\bm p_\lambda)
=\delta^3(\bm P-\bm P_c)\sum_m \langle l_\rho,m;l_\lambda,L_z-m|L,L_z \rangle
\psi^{\alpha_\rho}_{n_\rho l_\rho m }(\bm p_\rho)
\psi^{\alpha_\lambda}_{n_\lambda l_\lambda L_z-m }(\bm p_\lambda),
\end{align}
where
\begin{align}
\psi^\alpha_{nlm}(\bm p)=(i)^l(-1)^n \left[\frac{2n!}{(n+l+1/2)!} \right]^{1/2}\frac{1}{\alpha^{l+3/2}}\exp\left({-\frac{\bm p^2}{2\alpha^2}}\right)L_n^{l+1/2}(\bm p^2/\alpha^2)
\mathcal{Y}_{lm}(\bm p).
\end{align}
Note that the above wave functions are general forms for a 3-body system with arbitrary masses.

\section{The total wave function of the singly heavy baryons}
\label{app:wavefunction}

The ground states of the singly heavy baryons with $\frac{1}{2}^+$ belong to the mixed-symmetric 20 states of the SU(4) group. Since the flavor symmetry is badly broken, a more sensible way to classify the singly heavy baryons is by its light degrees of freedom. In the flavor space, they can be categorized by the light flavors into anti-triplet and sextet states according to $\text{SU(3)}\otimes \text{SU(3)}=\bar 3 \oplus 6$. For the convenience of constructing the total wave functions as representations of 3-dimension permutation group the heavy quark is usually labelled as the third quark. Then, the flavor wave functions of the anti-triplet and sextet are written as~\cite{Zhong:2007gp,Wang:2017kfr}:
\begin{align}
\phi^{\bar 3}_Q=
\begin{dcases}
\frac{1}{\sqrt2}(ud-du)Q, &~ \text{for}~ \Lambda_Q,  \\  
\frac{1}{\sqrt2}(us-su)Q, &~ \text{for}~ \Xi_Q,\\  
\frac{1}{\sqrt2}(ds-sd)Q, &~ \text{for}~ \Xi_Q,    
\end{dcases}
\end{align}
and
\begin{align}
\phi^{6}_Q=
\begin{dcases}
uuQ, &~ \text{for}~ \Sigma_Q,  \\  
\frac{1}{\sqrt2}(ud+du)Q, &~ \text{for}~ \Sigma_Q,\\  
ddQ, &~ \text{for}~ \Sigma_Q, \\
\frac{1}{\sqrt2}(us+su)Q, &~ \text{for}~ \Xi'_Q,\\ 
\frac{1}{\sqrt2}(ds+sd)Q, &~ \text{for}~ \Xi'_Q,\\
ssQ, &~ \text{for}~ \Omega_Q,       
\end{dcases}
\end{align}

With the anti-symmetric color wave function universal for the three-quark baryon systems, the total wave function, as the direct product of the spin, flavor and spatial wave functions, should keep symmetric in the light quark sector. For the ground states and the first orbital excitation states of the anti-triplet and sextet the wave functions are as follows:
\begin{align}
\begin{dcases}
\left |^2 S\frac{1}{2}^+ \right \rangle &=\Psi^s_{00}\chi_{S_z}^\rho\phi^{\bar 3}_Q \\
\left |^2 P_\lambda\frac{1}{2}^-\right\rangle &=\Psi^\lambda_{1L_z}\chi_{S_z}^\rho\phi^{\bar 3}_Q \\
\left |^2 P_\rho\frac{1}{2}^-\right\rangle &=\Psi^\rho_{1L_z}\chi_{S_z}^\lambda\phi^{\bar 3}_Q \\
\left |^4 P_\rho\frac{1}{2}^-\right\rangle &=\Psi^\rho_{1L_z}\chi_{S_z}^s\phi^{\bar 3}_Q
\end{dcases},
\label{eq:antitriplet}
\end{align}
for the anti-triplet states and 

\begin{align}
\begin{dcases}
\left|^2 S\frac{1}{2}^+\right\rangle &=\Psi^s_{00}\chi_{S_z}^\lambda\phi_B \\
\left|^2 P_\lambda\frac{1}{2}^-\right\rangle &=\Psi^\lambda_{1L_z}\chi_{S_z}^\lambda\phi_B \\
\left|^2 P_\rho\frac{1}{2}^-\right\rangle &=\Psi^\rho_{1L_z}\chi_{S_z}^\rho\phi_B \\
\left|^4 P_\lambda\frac{1}{2}^-\right\rangle &=\Psi^\lambda_{1L_z}\chi_{S_z}^s\phi_B
\end{dcases},
\label{eq:sextet}
\end{align}
for the sextet states. In the above two equations $\Psi$ is the spatial wave function given in Appendix~\ref{App:wavefuncion}. 

In the calculation we express the mesons and the singly heavy baryons with mock states~\cite{Hayne:1981zy}, i.e.,
\begin{eqnarray}\label{mock-meson}
| M(\bm P_c)_{J,J_z}\rangle &=&\sum_{S_z,L_z;c_i}\langle L,L_z;S,S_z |J,J_z \rangle\int d \bm p_1 d \bm p_2 \delta^3(\bm p_1 +\bm p_2-\bm P_c)\Psi_{N,L,L_z}(\bm p_1 ,\bm p_2)\chi_{s_1,s_2}^{S,S_z} \nonumber\\
&&\times\frac{\delta_{c_1 c_2}}{\sqrt3} \phi_{i_1,i_2}b^\dagger_{c_1,i_1,s_1,\bm p_1}d^\dagger_{c_2,i_2,s_2,\bm p_2}|0\rangle, 
\end{eqnarray}
for mesons, and
\begin{eqnarray}\label{mock-baryon}
| B(\bm P_c)_{J,J_z}\rangle &=&\sum_{S_z,L_z;c_i}\langle L,L_z;S,S_z |J,J_z \rangle\int d \bm p_1 d \bm p_2 d\bm p_3 \delta^3(\bm p_1 +\bm p_2+\bm p_3-\bm P_c) \Psi_{N,L,L_z}(\bm p_1 ,\bm p_2,\bm p_3) \chi_{s_1,s_2,s_3}^{S,S_z} \nonumber\\
&&\times\frac{\epsilon_{c_1 c_2 c_3}}{\sqrt6} \phi_{i_1,i_2,i_3} b^\dagger_{c_1,i_1,s_1,\bm p_1}b^\dagger_{c_2,i_2,s_2,\bm p_2}b^\dagger_{c_3,i_3,s_3,\bm p_3}|0\rangle,
\end{eqnarray}
for baryons. In the above two equations $c_j,\,s_j,\,i_j$ are color, spin, and flavor indices, respectively; $\psi_{N,L,L_z}$ denotes the spatial wave functions; $\chi^{S,S_z}$ is the spin wave function; $\phi$ is the flavor wave function, and $\delta_{c_1c_2}/\sqrt 3$ and $\epsilon_{c_1c_2c_3}/\sqrt6$ are the color wave functions for the meson and baryon, respectively; $\bm p_i$ denotes the single quark (antiquark) three-vector momentum, and $\bm P_c$ denotes the total momentum of the hadron (meson or baryon). The following normalization condition are used in this work:
\begin{equation}
\begin{aligned}
\langle  M(\bm P'_c)_{J,J_z} | M(\bm P_c)_{J,J_z}\rangle &=\delta^3(\bm P'_c-\bm P_c),\\
\langle  B(\bm P'_c)_{J,J_z} | B(\bm P_c)_{J,J_z}\rangle &=\delta^3(\bm P'_c-\bm P_c).
\label{baryon-norm}
\end{aligned}
\end{equation}

\section{amplitudes}\label{app:amplitudes}

The transition amplitudes $M^{J_f,J^z_f;J_i,J^z_i}$ satisfy the following relations:
\begin{eqnarray}
\mathcal M_{PC}^{1/2,1/2;1/2,1/2}&=&-\mathcal M_{PC}^{1/2,-1/2;1/2,-1/2},\\
\mathcal M_{PV}^{1/2,1/2;1/2,1/2}&=&\mathcal M_{PV}^{1/2,-1/2;1/2,-1/2}, \\ 
\mathcal M^{1/2,1/2;1/2,-1/2}_{PC/PV}&=&M^{1/2,-1/2;1/2,1/2}_{PC/PV}=0.
\end{eqnarray}
These relations can be easily verified by the amplitudes at the hadron level in the non-relativistic limit. As follows, only the amplitudes of $\mathcal M^{1/2,-1/2;1/2,-1/2}$ are provided and the spin index are omitted just for convenience. 
We also note that a relative phase difference of $\pi$ between the pole terms and the CS or DPE processes is implied. It arises from coupling vertices and propagators.

All the amplitudes for the processes that heavy quark acts as a spectator can be related by flavor symmetry. As for the $\Xi_c$ decay sector, the $cs\to dc$ weak transition processes are also related by flavor symmetry. So only the amplitudes of $\Xi_c^0\to\Lambda_c \pi^-$ are provided here. 

Since those processes with the heavy quark as the spectator have the same expressions for both $\Xi_c$ and $\Xi_b$ decays, we only provide  the amplitudes of $\Xi_c^0\to\Lambda_c \pi^-$. The amplitudes of the other three channels can be obtained by proper substitutions of the corresponding quantities or parameters.


Here we first define several functions which will appear later in the amplitudes:
\begin{align}
\xi_A&=\exp  \left[-\frac{\omega_k^2}{8}  \left(\frac{1}{\alpha_{\rho,c}^2}\frac{4m_s^2}{(m_s+m_q)^2}+\frac{1}{\alpha _{\lambda,c}^2}\frac{3 m_c^2}{\left(m_c+m_q+m_s\right){}^2}\right)\right],\\
\xi_B&=\exp \left [ -\frac{3\omega_k^2}{4}\frac{(m_q-m_s)^2m_c^2}{(m_c+2m_q)^2(m_c+m_s+m_q)^2(\alpha_{\lambda,c}^2+\alpha_{\lambda,sc}^2)}
 \right],\\
\xi_C&=\exp\left[
-\frac{3m_q^2\omega_k^2(m_q^2-m_s^2)^2}{(m_c+2m_q)^2(m_q+m_s+m_c)^2
[(m_q+m_s)^2(\alpha_{\lambda,c}^2+3(\alpha_{\rho,c}^2+\alpha_{\rho,sc}^2))+4m_q^2\alpha_{\lambda,sc}^2]}
 \right],
\end{align}
The harmonic oscillator strength defined as:
\begin{align}
&\alpha_{\rho,c/b}=\left(3 K_{{c/b}} m_q\right)^{1/4}, 
&& \alpha_{\lambda,c/b}=\left(3 K_{c/b} \frac{3 m_q m_{c/b}}{2m_q+m_{c/b}}\right)^{1/4},\notag \\
&\alpha_{\rho,sc/sb}=\left(3 K_{c/b} \frac{2m_q m_s}{m_q+m_s}\right)^{1/4}, 
&& \alpha_{\lambda,sc/sb}=\left(3 K_{c/b} \frac{2m_{c/b}(m_q^2+m_s^2+m_q m_s)}{(m_q+m_s+m_{c/b})(m_q+m_s)}\right)^{1/4}.
\end{align}

The propagator is expressed with
\begin{align}
\mathcal{P}(B_i,B_f)=\frac{2m_{B_f}}{m_{B_i}^2-m_{B_f}^2+i \Gamma_{B_f} m_{B_f}}.
\end{align}

\subsection{$\Xi_c^0\to\Lambda_c \pi^-$}

\begin{itemize}
\item Pole terms ($cs\to dc$)
\begin{align}
&\mathcal{M}_{\text{Pole-A}, PC}^{cs\to dc}(\Sigma_c^0)_{\Xi_c^0\to\Lambda_c \pi^-}\notag \\
=&\frac{8\sqrt 3 G_F V_{cd}V_{cs}}{\pi^{3/2}}(m_q+m_s)^3
\left[\frac{\alpha_{\lambda,c}\alpha_{\lambda,sc}\alpha_{\rho,c}\alpha_{\rho,sc}}{(m_q+m_s)^2(\alpha_{\lambda,c}^2+3(\alpha_{\rho,c}^2+\alpha_{\rho,sc}^2))+4m_q^2\alpha_{\lambda,sc}^2}\right]^{3/2}  \notag \\
\times&
\mathcal{P}(\Xi_c^0,\Sigma_c^0)\times\frac{\omega_k(\omega_0+2m_c+4m_q)}{4\sqrt 6\pi^{3/2}(m_c+2m_q)f_\pi\sqrt{\omega_0}}
\xi'_{A},
\end{align}
where
$\xi'_{A}$ means $m_s$ appears in $\xi_{A}$ should be replaced by $m_q$.

\begin{align}
&\mathcal{M}_{\text{Pole-A}, PV}^{cs\to dc}(\Sigma_c^0|^2P_\lambda \rangle)_{\Xi_c^0\to\Lambda_c \pi^-}\notag \\
=&i\frac{\omega_k^2m_cm_q(\omega_0+ 2m_c+4m_q)-2\omega_0\alpha_{\lambda,c}^2(m_c+2m_q)^2}{8\sqrt6 \pi^{3/2} \alpha_{\lambda,c}m_q(m_c+2m_q)^2 f_ \pi\sqrt{\omega_0}} \xi'_A
 \times  \mathcal{P}(\Xi_c^0,\Sigma_c|^2 P_\rho\rangle) 
 \notag \\
 \times& i\frac{4 G_FV_{cs}V_{cd}}{\sqrt3 m_q m_sm_c\pi^{3/2}}
 \frac{(m_q+m_s)^3\alpha_{\lambda,c}(\alpha_{\lambda,c}\alpha_{\lambda,sc}\alpha_{\rho,c}\alpha_{\rho,sc})^{3/2}}{\left[(m_q+m_s)^2(\alpha_{\lambda,c}^2+3(\alpha_{\rho,c}^2+\alpha_{\rho,sc}^2))+4m_q^2\alpha_{\lambda,sc}^2\right]^{5/2}}
 \notag \\
 \times& \left[
4\alpha_{\lambda,sc}^2 m_q^2m_s(2m_c+3m_s+m_q)
-6\alpha_{\rho,c}^2m_s(m_c+m_q)(m_s+m_q)^2
\right. \notag \\ 
&\left.
-3\alpha_{\rho,sc}^2(m_q+m_s)^2(m_c m_s+2 m_q m_s+3m_q m_c)
 \right]
\end{align}

\begin{align}
&\mathcal{M}_{\text{Pole-A}, PV}^{cs\to dc}(\Sigma_c^0|^4P_\rho \rangle)_{\Xi_c^0\to\Lambda_c \pi^-}
\notag \\
=&i\frac{\omega_k^2m_cm_q(\omega_0+2m_c+4m_q)-\omega_0\alpha_{\lambda,c}^2(m_c+2m_q)^2+}{8\sqrt3 \pi^{3/2} \alpha_{\lambda,c}m_q(m_c+2m_q)^2 f_\pi \sqrt{\omega_0}} \xi'_A
 \times  \mathcal{P}(\Xi_c^0,\Sigma_c|^2 P_\rho\rangle) 
 \notag \\
 \times& i \frac{8 G_FV_{cs}V_{cd}}{\sqrt6 m_q m_c\pi^{3/2}}
 \frac{(m_q+m_s)^3\alpha_{\lambda,c}(\alpha_{\lambda,c}\alpha_{\lambda,sc}\alpha_{\rho,c}\alpha_{\rho,sc})^{3/2}}{\left[(m_q+m_s)^2(\alpha_{\lambda,c}^2+3(\alpha_{\rho,c}^2+\alpha_{\rho,sc}^2))+4m_q^2\alpha_{\lambda,sc}^2\right]^{5/2}}
 \notag \\
 \times& \left\{
 (2m_q+m_c)\left[3(m_q+m_s)^2(\alpha_{\rho,c}^2+\alpha_{\rho,sc}^2)+4m_q^2\alpha_{\lambda,sc}^2\right]+3m_c(m_q+m_s)^2\alpha_{\lambda,sc}^2
 \right\}
\end{align}

\begin{align}
&\mathcal{M}_{\text{Pole-B}, PC}^{cs\to dc}(\Xi'^+_c)_{\Xi_c^0\to\Lambda_c \pi^-}
\notag \\
=&\frac{(2m_q+2m_s+2m_c+\omega_0)\omega_k}{8\sqrt 3 \pi^{3/2}f_\pi \sqrt{\omega_0}(m_q+m_s+m_c)}\xi_A \times \mathcal{P}(\Lambda_c,\Xi'^+_c)
\notag \\
\times& 
\frac{4\sqrt 6 G_F V_{cd}V_{cs}}{\pi^{3/2}}
(m_s+m_q)^3
\left[\frac{\alpha_{\lambda,c}\alpha_{\lambda,sc}\alpha_{\rho,c}\alpha_{\rho,sc}}{(m_q+m_s)^2(\alpha_{\lambda,c}^2+3(\alpha_{\rho,c}^2+\alpha_{\rho,sc}^2))+4m_q^2\alpha_{\lambda,sc}^2}\right]^{3/2}\times\xi_C,
\end{align}

\begin{align}
&\mathcal{M}_{\text{Pole-B}, PV}^{cs\to dc}(\Xi^+_c|^2P_\rho \rangle)_{\Xi_c^0\to\Lambda_c \pi^-} \notag  \\
=&i\frac{[3\omega_0\alpha_{\rho,sc}^2(m_q+m_s)+2\omega_k^2 m_qm_s](m_q+m_s+m_c)+m_qm_s\omega_0\omega_k^2}{24\pi^{3/2}f_\pi \sqrt{\omega_0}\alpha_{\rho,sc}m_q(m_s+m_q)(m_c+m_s+m_q)}\xi_A \times \mathcal{P}(\Lambda_c,\Xi_c^+|^2P_\rho)
\notag  \\
\times & \frac{-2\sqrt{2}i G_F V_{cd} V_{cs}}{3\pi^{3/2}}(m_q+m_s)^4
\frac{\alpha_{\rho,sc}(\alpha_{\lambda,c}\alpha_{\lambda,sc}\alpha_{\rho,c}\alpha_{\rho,sc})^{3/2} }{m_qm_sm_c} \times \xi_C
\notag  \\
\times &
\left[ 
\frac{2[4\alpha_{\lambda,sc}^2m_q^2(m_s+m_c)+3\alpha_{\rho,c}^2m_c(m_q+m_s)(m_q-3m_s)+\alpha_{\lambda,c}^2(m_q+m_s)(6m_qm_s+m_qm_c+3m_sm_c)]}{[(m_q+m_s)^2(\alpha_{\lambda,c}^2+3(\alpha_{\rho,c}^2+\alpha_{\rho,sc}^2))+4m_q^2\alpha_{\lambda,sc}^2]^{5/2}}
 \right.
\notag \\
+&\left.
\frac{\zeta_1}{(m_c+2m_q)^2(m_q+m_s+m_c)^2
[(m_q+m_s)^2(\alpha_{\lambda,c}^2+3(\alpha_{\rho,c}^2+\alpha_{\rho,sc}^2))+4m_q^2\alpha_{\lambda,sc}^2]^{7/2}}
\right],
\end{align}
where
\begin{align}
\zeta_1
&=18(m_q-m_s)^2(m_s+m_q)^3\omega_k^2
\left[\alpha_{\rho,sc}^2m_cm_q^3+3\alpha^2_{\rho,c}m_cm_sm_q^2-\alpha^2_{\lambda,c}m_s(m_c+m_q)m_q^2
\right]
 \notag \\
&-24\alpha^2_{\lambda,sc}\omega_k^2 m_s m_q^4(m_q-m_s)^2(m_q+m_s)(m_q+m_s+m_c)
+(m_c+2m_q)^2(m_q+m_s+m_c)^2
\notag \\
&\times \left\{
(m_q+m_s)^3 \left[ \alpha_{\lambda,c}^4(3m_sm_c+6m_qm_s+m_cm_q)+9\alpha_{\rho,c}^4m_c(m_q-3m_s) \right]+16\alpha_{\lambda,sc}^4(m_c+m_s)m_q^4 
\right .\notag\\
&+4m_q^2(m_s+m_q)\alpha_{\lambda,sc}^2 \left[ \alpha_{\lambda,c}^2(4m_sm_c+2m_qm_c+7m_qm_s+m_s^2)
+3\alpha_{\rho,c}^2(2m_qm_c+m_sm_q+m_s^2-2m_sm_c) \right]
\notag\\
&+3(m_q+m_s)^3\alpha_{\rho,sc}^2 \left[ 4\alpha_{\lambda,sc}^2m_q^2(m_s+m_c)+\alpha_{\lambda,c}^2(m_s+m_q)(3m_sm_c+6m_qm_s+m_qm_c) \right]
\notag \\
&\left . +3(m_q+m_s)^3\alpha_{\rho,c}^2 \left[ 2 \alpha_{\lambda,c}^2m_q(3m_s+m_c)+3 \alpha_{\rho,sc}^2m_c(m_q-3m_s)^2 \right]
 \right\}.
\end{align}

\begin{align}
&\mathcal{M}_{\text{Pole-B}, PV}^{cs\to dc}(\Xi^+_c|^4P_\rho \rangle)_{\Xi_c^0\to\Lambda_c \pi^-}
\notag \\
=&i\frac{[3\omega_0\alpha_{\rho,sc}^2(m_q+m_s)+4\omega_k^2 m_qm_s](m_q+m_s+m_c)+2m_qm_s\omega_0\omega_k^2}{24\sqrt{2}\pi^{3/2}f_\pi \sqrt{\omega_0}\alpha_{\rho,sc}m_q(m_s+m_q)(m_c+m_s+m_q)}\xi_A \times \mathcal{P}(\Lambda_c,\Xi_c^+|^4P_\rho)
\notag  \\
\times & \frac{2i G_F V_{cd} V_{cs}}{3\pi^{3/2}}(m_q+m_s)^4
\frac{\alpha_{\rho,sc}(\alpha_{\lambda,c}\alpha_{\lambda,sc}\alpha_{\rho,c}\alpha_{\rho,sc})^{3/2} }{m_qm_sm_c} \times \xi_C
\notag\\
\times &
\left[ 
\frac{2m_qm_c(m_q+m_s)\alpha_{\lambda,c}^2+2m_q^2(m_s+m_c)(4\alpha_{\lambda,sc}^2+3\alpha_{\rho,c}^2)}{[(m_q+m_s)^2(\alpha_{\lambda,c}^2+3(\alpha_{\rho,c}^2+\alpha_{\rho,sc}^2))+4m_q^2\alpha_{\lambda,sc}^2]^{5/2}}
 \right.
\notag \\
+&\left.
\frac{\zeta_2}{(m_c+2m_q)^2(m_q+m_s+m_c)^2
[(m_q+m_s)^2(\alpha_{\lambda,c}^2+3(\alpha_{\rho,c}^2+\alpha_{\rho,sc}^2))+4m_q^2\alpha_{\lambda,sc}^2]^{7/2}}
\right],
\end{align}
where
\begin{align}
\zeta_2
&=24(m_q+m_s)(m_q-m_s)^2m_q^3\omega_k^2
\left[3\alpha_{\rho,sc}^2m_c(m_q+m_s)^2-4\alpha_{\lambda,sc}^2m_qm_s(m_q+m_s+m_c)
\right] \notag \\
&+(2m_q+m_c)^2(m_q+m_s+m_c)^2 
\left\{ 16m_q^2(m_s+m_c)(4 \alpha_{\lambda,sc}^4 m_q^2+3\alpha_{\lambda,sc}^2\alpha_{\rho,sc}^2(m_q+m_s)^2)
\right . \notag \\
&+4m_qm_c(m_q+m_s)^3(\alpha_{\lambda,c}^4+9\alpha_{\rho,c}^4+3\alpha_{\lambda,c}^2\alpha_{\rho,sc}^2+6\alpha_{\lambda,c}^2\alpha_{\rho,c}^2+9\alpha_{\rho,c}^2\alpha_{\rho,sc}^2)
\notag \\
&\left.+16(m_s+m_q)m_q^2\alpha_{\lambda,sc}^2(\alpha_{\lambda,c}^2+3\alpha_{\rho,c}^2)(m_s^2+m_sm_q+2m_qm_c+m_sm_c)\right\}.
\end{align}

\begin{align}
&\mathcal{M}_{\text{Pole-B}, PV}^{cs\to dc}(\Xi^{'+}_c|^2P_\lambda \rangle)_{\Xi_c^0\to\Lambda_c \pi^-}
\notag \\
=& i\frac{4\omega_0\alpha_{\lambda,sc}^2(m_q+m_s+m_c)^2+\omega_k^2m_c(m_q+m_s)(2m_q+2m_s+2m_c+\omega_0)}{16\sqrt{3}\pi^{3/2}f_\pi \sqrt{\omega_0}\alpha_{\lambda,sc}(m_s+m_q)(m_c+m_s+m_q)^2}\xi_A \times \mathcal{P}(\Lambda_c,\Xi_c^{'+}|^2P_\rho)
\notag  \\
\times& \frac{8i G_F V_{cd} V_{cs}}{3\sqrt{6}\pi^{3/2}}(m_q+m_s)^4
\frac{\alpha_{\lambda,sc}(\alpha_{\lambda,c}\alpha_{\lambda,sc}\alpha_{\rho,c}\alpha_{\rho,sc})^{3/2} }{m_qm_s} \times \xi_C
\notag\\
\times &\left[ 2\frac{\alpha_{\lambda,c}^2m_s(m_s-2m_c-5m_q)+3\alpha_{\rho,c}^2m_s(m_q+m_s+4m_c)+3\alpha_{\rho,sc}(m_s+m_c)(m_q+m_s)}{[(m_q+m_s)^2(\alpha_{\lambda,c}^2+3(\alpha_{\rho,c}^2+\alpha_{\rho,sc}^2))+4m_q^2\alpha_{\lambda,sc}^2]^{5/2}} \right.
\notag \\
+&\left.
\frac{\zeta_3}{(m_c+2m_q)^2(m_q+m_s+m_c)^2
[(m_q+m_s)^2(\alpha_{\lambda,c}^2+3(\alpha_{\rho,c}^2+\alpha_{\rho,sc}^2))+4m_q^2\alpha_{\lambda,sc}^2]^{7/2}}
\right],
\end{align}
where
\begin{align}
\zeta_3&=
18m_q^2(m_q-m_s)^2\omega_k^2
\left\{(m_q+m_s)^2\left[\alpha_{\lambda,c}^2m_s(2m_q+m_c)-\alpha_{\rho,sc}^2m_qm_c-3\alpha_{\rho,c}^2m_sm_c
\right]
\right.
\notag \\
&+3\alpha^2m_sm_q^2(m_q+m_s+m_c) \left. \right\} +(m_c+2m_q)^2(m_q+m_s+m_c)^2
\notag \\
&\times \left\{(m_s+m_q)^2\left[
\alpha_{\lambda,c}^4m_s(m_s-5m_q-2m_c)+9\alpha_{\rho,sc}^4(m_q+m_s)(m_s+m_c)
\right.  \right.\notag \\
&+9\alpha_{\lambda,sc}^4m_s(m_q+m_s+4m_c)+3\alpha_{\lambda,c}^2\alpha_{\rho,sc}^2(m_c(m_q-m_s)+2m_s(m_s-2m_q))
\notag \\
&+6\alpha_{\lambda,c}^2\alpha_{\rho,c}^2m_s(m_c+m_s-2m_q)
+\left. 9\alpha_{\rho,c}^2\alpha_{\rho,sc}^2(m_c(m_q+5m_s)+2m_s(m_q+m_s)) \right]
\notag \\
&+4m_q^2 \left[
\alpha_{\lambda,c}^2\alpha_{\lambda,sc}^2m_s(m_s-5m_q-2m_c)+3\alpha_{\lambda,sc}^2\alpha_{\rho,sc}^2(m_s+m_c)(m_q+m_s) \right.
\notag \\
&\left. \left.+3\alpha_{\lambda,sc}^2\alpha_{\rho,c}^2m_s(m_q+m_s+4m_c)
\right] \right\}
\end{align}

\begin{align}
&\mathcal{M}_{\text{Pole-B}, PV}^{cs\to dc}(\Xi'^+_c|^4P_\rho \rangle)_{\Xi_c^0\to\Lambda_c \pi^-}
\notag \\
=& i\frac{2\omega_0\alpha_{\lambda,sc}^2(m_q+m_s+m_c)^2+\omega_k^2m_c(m_q+m_s)(2m_q+2m_s+2m_c+\omega_0)}{8\sqrt{6}\pi^{3/2}f_\pi \sqrt{\omega_0}\alpha_{\lambda,sc}(m_s+m_q)(m_c+m_s+m_q)^2}\xi_A \times \mathcal{P}(\Lambda_c,\Xi_c^{'+}|^4P_\rho)
\notag \\
\times& \frac{-4i G_F V_{cd} V_{cs}}{3\sqrt{3}\pi^{3/2}}(m_q+m_s)^4
\frac{\alpha_{\lambda,sc}(\alpha_{\lambda,c}\alpha_{\lambda,sc}\alpha_{\rho,c}\alpha_{\rho,sc})^{3/2} }{m_qm_s} \times \xi_C
\notag\\
\times &\left[ 2\frac{(\alpha_{\lambda,c}^2+3\alpha_{\rho,c}^2)m_s(m_q+m_s+m_c)^2+3\alpha_{\rho,sc}^2(m_s+m_c)(m_q+m_s)}{[(m_q+m_s)^2(\alpha_{\lambda,c}^2+3(\alpha_{\rho,c}^2+\alpha_{\rho,sc}^2))+4m_q^2\alpha_{\lambda,sc}^2]^{5/2}} \right.
\notag \\
+&\left.
\frac{4\zeta_4}{(m_c+2m_q)^2(m_q+m_s+m_c)^2
[(m_q+m_s)^2(\alpha_{\lambda,c}^2+3(\alpha_{\rho,c}^2+\alpha_{\rho,sc}^2))+4m_q^2\alpha_{\lambda,sc}^2]^{7/2}}
\right],
\end{align}
where
\begin{align}
\zeta_4&= 6\omega_k^2(m_q-m_s)^2m_q^3 \left[ 4 \alpha_{\lambda,sc}^2m_qm_s(m_q+m_s+m_c)-3\alpha_{\rho,sc}^2m_c(m_q+m_s)^2 \right] 
\notag \\
&+(2m_q+m_c)^2(m_q+m_s+m_c)^2
\left\{ m_s(m_q+m_s)^2(m_q+m_s+m_c) (\alpha_{\lambda,c}^2+3\alpha_{\rho,c}^2)^2 \right.
\notag \\
&+3\alpha_{\rho,sc}^2(\alpha_{\lambda,c}^2+3\alpha_{\rho,c}^2)(m_q+m_s)^2\left[(m_q+m_s)(m_c+2m_s)+m_sm_c\right]
\notag \\
&+3(m_s+m_c)(m_q+m_s)\alpha_{\rho,sc}^2[4m_q^2\alpha_{\lambda,sc}+3\alpha_{\rho,sc}^2(m_q+m_s)^2]
\notag \\
&\left. +4m_sm_q^2(m_q+m_s+m_c) \left[\alpha_{\lambda,c}^2\alpha_{\lambda,sc}^2+3\alpha_{\rho,c}^2\alpha_{\rho,sc}^2\right] \right \} .
\end{align}

\item Pole terms ($us\to du$)
\begin{align}
&\mathcal{M}_{\text{Pole-B}, PV}^{us\to du}(\Xi_c^+|^2 P_\rho\rangle)_{\Xi_c^0\to\Lambda_c \pi^-}\notag \\
=&\left[
i\frac{\left(2\omega_k^2 m_q m_s+3 \omega_0(m_s+m_q)\alpha_{\rho,sc}^2\right)(m_q+m_s+m_c)+m_qm_s \omega_0\omega_k^2}{24\pi^{3/2} f_\pi \sqrt{\omega_0}m_q(m_s+m_q)(m_q+m_s+m_c) \alpha_{\rho,sc}}
\right] \times \xi_A
\times \mathcal{P}(\Xi_c^0,\Xi_c^+) \notag \\
\times&
\frac{-i\sqrt2 G_FV_{ud}V_{us}\alpha_{\rho,sc}}{\pi^{3/2}}
\frac{m_q+m_s}{m_q m_s}
\left[ \frac{\alpha_{\lambda,c}\alpha_{\lambda,sc}\alpha_{\rho,c}\alpha_{\rho,sc}}{\alpha_{\lambda,c}^2+\alpha_{\lambda,sc}^2}\right]^{3/2} \times \xi_B
\end{align}

\begin{align}
&\mathcal{M}_{\text{Pole-B}, PV}^{us\to du}(\Xi_c^+|^4 P_\rho\rangle)_{\Xi_c^0\to\Lambda_c \pi^-} \notag \\
=&\left[
i\frac{\left(4\omega_k^2 m_q m_s+ 3\omega_0(m_s+m_q)\alpha_{\rho,sc}^2\right)(m_q+m_s+m_c)+2m_q \omega_0\omega_k^2}{24\sqrt2\pi^{3/2} f_\pi \sqrt{\omega_0}m_q(m_s+m_q)(m_q+m_s+m_c) \alpha_{\rho,sc}}
\right] \times \xi_A \times  \mathcal{P}(\Xi_c^0,\Xi_c^+)\notag \\
\times& \frac{-iG_FV_{ud}V_{us}\alpha_{\rho,sc}}{\pi^{3/2}}
\frac{m_q+m_s}{m_q m_s}
\left[ \frac{\alpha_{\lambda,c}\alpha_{\lambda,sc}\alpha_{\rho,c}\alpha_{\rho,sc}}{\alpha_{\lambda,c}^2+\alpha_{\lambda,sc}^2}\right]^{3/2} \times \xi_B
\end{align}

\item CS
\begin{align}
&\mathcal{M}_{\text{CS},PV}(\Xi_c^0\to\Lambda_c \pi^-) \notag  \\
=&-\frac{\sqrt 3 G_F V_{ud}V_{us}\alpha_{\rho,c}^{3/2} }{4 (\pi^{3/2} \alpha_{\lambda,c} \alpha_{\lambda,sc}\alpha_{\rho,sc} R)^{3/2}}
\left[(\frac{1}{\alpha_{\rho,sc}^2}+\frac{1}{2R^2})
(\frac{3}{4 \alpha_{\lambda,c}^2} +\frac{3}{4 \alpha_{\lambda,sc}^2} 
+\frac{m_q^2}{(m_q+m_s)^2(\alpha^2_{\rho,sc}+2R^2)})  \right]^{-3/2} \notag \\
\times & \exp(-3\omega_k^2\frac{N_{CS}}{D_{CS}}),
\end{align}
where
\begin{align}
N_{CS}&=(m_c+2m_q)^2(m_q+m_s)^2(\alpha_{\lambda,c}^2+\alpha_{\lambda,sc}^2) 
+3m_c^2(m_q+m_s)^2(\alpha_{\rho,sc}^2+2R^2)\notag \\
&-4m_cm_q\alpha_{\rho,sc}^2 (2m_q^2+2m_qm_s+m_cm_s), \\
D_{CS}&=4(m_c+m_q)^2 \left[3(m_q+m_s)^2(\alpha_{\lambda,c}^2+\alpha_{\lambda,sc}^2)(\alpha_{\rho,sc}^2+2R^2)+4m_q^2 \alpha_{\lambda,c}^2 \alpha_{\lambda,sc}^2\right].
\end{align}

\item DPE
\begin{align}
&\mathcal{M}_{\text{DPE},PV}(\Xi_c^0\to\Lambda_c \pi^-)\notag \\
=&\frac{6\sqrt 6 G_F V_{ud} V_{us}}{\pi^{9/4}}(m_q+m_s)^3 \alpha^{1/2}_{\rho,sc}
(\alpha_{\lambda,c}\alpha_{\lambda,sc}\alpha_{\rho,c}R) ^{3/2} 
\notag \\
\times& \left[3(m_q+m_s)^2 \alpha_{\rho,sc}^2(\alpha_{\lambda,c}^2+\alpha_{\lambda,sc}^2)+(m_q-m_s)^2 \alpha_{\lambda,c}^2\alpha_{\lambda,sc}^2\right]^{-1/2}
\notag \\
\times& \left[3(m_q+m_s)^2 (\alpha_{\rho,c}^2+\alpha_{\rho,sc}^2)(\alpha_{\lambda,c}^2+\alpha_{\lambda,sc}^2)+(m_q-m_s)^2 \alpha_{\lambda,c}^2\alpha_{\lambda,sc}^2\right]^{-1} 
\notag \\
\times& \left[\frac{1}{\alpha_{\rho,c}^2}+\frac{1}{\alpha_{\rho,sc}^2} -\frac{(m_q-m_s)^2}{\alpha_{\rho,sc}^2(m_q+m_s)^2 }\left(\frac{3\alpha_{\rho,sc}^2}{\alpha_{\lambda,c}^2}+\frac{3\alpha_{\rho,sc}^2}{\alpha_{\lambda,sc}^2}+(\frac{m_q-m_s}{m_q+m_s})^2\right)^{-1}\right]^{-1/2} 
\notag \\
\times&\exp\left[-\frac{3}{4}\omega_k^2
\frac{N_{DPE}}{D_{DPE}} \right],
\end{align}
where
\begin{align}
N_{DPE}&=(m_c+2m_q)^2(m_q+m_s)^2\alpha_{\lambda,c}^2+3m_c^2(m_q+m_s)^2(\alpha_{\rho,c}^2+\alpha_{\rho,sc}^2)+4m_q^2(m_c+m_q)^2\alpha_{\lambda,sc}^2) \notag \\
&+4m_q m_c\alpha_{\lambda,sc}^2(2m_sm_q+m_c m_q+2m_q^2+m_s^2), \\
D_{DPE}&=(m_c+2m_q)^2\left[3(m_q+m_s)^2(\alpha_{\lambda,c}^2+\alpha_{\lambda,sc}^2)(\alpha_{\rho,c}^2+\alpha_{\rho,sc}^2)
+(m_q-m_s)^2\alpha_{\lambda,c}^2\alpha_{\lambda,sc}^2\right].
\end{align}

\end{itemize}

\subsection{$\Xi_c^+\to\Lambda_c \pi^0$}
\begin{itemize}
\item Pole terms ($cs\to dc$)
\begin{align}
\mathcal{M}_{\text{Pole-A}, PC}^{cs\to dc}(\Sigma_c^+)_{\Xi_c^+\to\Lambda_c \pi^0}
=\frac{1}{\sqrt 2}
\mathcal{M}_{\text{Pole-A}, PC}^{cs\to dc}(\Sigma_c^0)_{\Xi_c^0\to\Lambda_c \pi^-}
\end{align}

\begin{align}
\mathcal{M}_{\text{Pole-A}, PV}^{cs\to dc}(\Sigma_c^+|^2P_\lambda \rangle)_{\Xi_c^+\to\Lambda_c \pi^0}
=\frac{1}{\sqrt 2}
\mathcal{M}_{\text{Pole-A}, PV}^{cs\to dc}(\Sigma_c^0|^2P_\lambda \rangle)_{\Xi_c^0\to\Lambda_c \pi^-}
\end{align}

\begin{align}
\mathcal{M}_{\text{Pole-A}, PV}^{cs\to dc}(\Sigma_c^+|^4P_\rho \rangle)_{\Xi_c^+\to\Lambda_c \pi^0}
=\frac{1}{\sqrt 2}
\mathcal{M}_{\text{Pole-A}, PV}^{cs\to dc}(\Sigma_c^0|^4P_\rho \rangle)_{\Xi_c^0\to\Lambda_c \pi^-}
\end{align}

\begin{align}
\mathcal{M}_{\text{Pole-B}, PC}^{cs\to dc}(\Xi'^+_c)_{\Xi_c^+\to\Lambda_c \pi^0}
=\frac{1}{\sqrt 2}
\mathcal{M}_{\text{Pole-B}, PC}^{cs\to dc}(\Xi'^+_c)_{\Xi_c^0\to\Lambda_c \pi^-}
\end{align}

\begin{align}
\mathcal{M}_{\text{Pole-B}, PV}^{cs\to dc}(\Xi^+_c|^2P_\rho \rangle)_{\Xi_c^+\to\Lambda_c \pi^0}
=\frac{1}{\sqrt 2}
\mathcal{M}_{\text{Pole-B}, PV}^{cs\to dc}(\Xi^+_c|^2P_\rho \rangle)_{\Xi_c^0\to\Lambda_c \pi^-}
\end{align}
\begin{align}
\mathcal{M}_{\text{Pole-B}, PV}^{cs\to dc}(\Xi^+_c|^4P_\rho \rangle)_{\Xi_c^+\to\Lambda_c \pi^0}
=\frac{1}{\sqrt 2}
\mathcal{M}_{\text{Pole-B}, PV}^{cs\to dc}(\Xi^+_c|^4P_\rho \rangle)_{\Xi_c^0\to\Lambda_c \pi^-}
\end{align}

\begin{align}
\mathcal{M}_{\text{Pole-B}, PV}^{cs\to dc}(\Xi'^+_c|^2P_\lambda \rangle)_{\Xi_c^+\to\Lambda_c \pi^0}
=\frac{1}{\sqrt 2}
\mathcal{M}_{\text{Pole-B}, PV}^{cs\to dc}(\Xi'^+_c|^2P_\lambda \rangle)_{\Xi_c^0\to\Lambda_c \pi^-}
\end{align}
\begin{align}
\mathcal{M}_{\text{Pole-B}, PV}^{cs\to dc}(\Xi'^+_c|^4P_\rho \rangle)_{\Xi_c^+\to\Lambda_c \pi^0}
=\frac{1}{\sqrt 2}
\mathcal{M}_{\text{Pole-B}, PV}^{cs\to dc}(\Xi'^+_c|^4P_\rho \rangle)_{\Xi_c^0\to\Lambda_c \pi^-}
\end{align}

\item Pole terms ($us\to du$)
\begin{align}
\mathcal{M}_{\text{Pole-B}, PV}^{us\to du}(\Xi_c^+|^2 P_\rho\rangle)_{\Xi_c^+\to\Lambda_c \pi^0}
=\frac{1}{\sqrt 2}
\mathcal{M}_{\text{Pole-B}, PV}^{us\to du}(\Xi_c^+|^2 P_\rho\rangle)_{\Xi_c^0\to\Lambda_c \pi^-}
\end{align}

\begin{align}
\mathcal{M}_{\text{Pole-B}, PV}^{us\to du}(\Xi_c^+|^4 P_\rho\rangle)_{\Xi_c^+\to\Lambda_c \pi^0}
=\frac{1}{\sqrt 2}
\mathcal{M}_{\text{Pole-B}, PV}^{us\to du}(\Xi_c^+|^4 P_\rho\rangle)_{\Xi_c^0\to\Lambda_c \pi^-}
\end{align}

\item CS
\begin{align}
\mathcal{M}_{\text{CS},PV}(\Xi_c^+\to\Lambda_c \pi^0)
=\frac{1}{\sqrt 2}
\mathcal{M}_{\text{CS},PV}(\Xi_c^0\to\Lambda_c \pi^-)
\end{align}

\item DPE
\begin{align}
\mathcal{M}_{\text{DPE},PV}(\Xi_c^+\to\Lambda_c \pi^0)=0
\end{align}

\end{itemize}

\subsection{$\Xi_b^-\to\Lambda_b \pi^-$}
\begin{itemize}
\item Pole terms ($us\to du$)
\begin{align}
\mathcal{M}_{\text{Pole-B}, PV}^{us\to du}(\Xi_b^0|^2 P_\rho\rangle)_{\Xi_b^-\to\Lambda_b \pi^-}
=
\mathcal{M}_{\text{Pole-B}, PV}^{us\to du}(\Xi_c^+|^2 P_\rho\rangle)_{\Xi_c^0\to\Lambda_c \pi^-}
\end{align}

\begin{align}
\mathcal{M}_{\text{Pole-B}, PV}^{us\to du}(\Xi_b^0|^4 P_\rho\rangle)_{\Xi_b^-\to\Lambda_b \pi^-}
=
\mathcal{M}_{\text{Pole-B}, PV}^{us\to du}(\Xi_c^+|^4 P_\rho\rangle)_{\Xi_c^0\to\Lambda_c \pi^-}
\end{align}

\item CS
\begin{align}
\mathcal{M}_{\text{CS},PV}(\Xi_b^-\to\Lambda_b \pi^-)
=
\mathcal{M}_{\text{CS},PV}(\Xi_c^0\to\Lambda_c \pi^-)
\end{align}

\item DPE
\begin{align}
\mathcal{M}_{\text{DPE},PV}(\Xi_b^-\to\Lambda_b \pi^-)
=
\mathcal{M}_{\text{DPE},PV}(\Xi_c^0\to\Lambda_c \pi^-)
\end{align}

\end{itemize}

\subsection{$\Xi_b^0\to\Lambda_b \pi^0$}
\begin{itemize}
\item Pole terms ($us\to du$)
\begin{align}
\mathcal{M}_{\text{Pole-B}, PV}^{us\to du}(\Xi_b^0|^2 P_\rho\rangle)_{\Xi_b^0\to\Lambda_b \pi^0}
=
\frac{1}{\sqrt 2}\mathcal{M}_{\text{Pole-B}, PV}^{us\to du}(\Xi_b^0|^2 P_\rho\rangle )_{\Xi_b^-\to\Lambda_b \pi^-}
\end{align}

\begin{align}
\mathcal{M}_{\text{Pole-B}, PV}^{us\to du}(\Xi_b^0|^4 P_\rho\rangle )_{\Xi_b^0\to\Lambda_b \pi^0}
=
\frac{1}{\sqrt 2}\mathcal{M}_{\text{Pole-B}, PV}^{us\to du}(\Xi_b^0|^4 P_\rho\rangle )_{\Xi_b^-\to\Lambda_b \pi^-}
\end{align}

\item CS
\begin{align}
\mathcal{M}_{\text{CS}, PV}(\Xi_b^0\to\Lambda_b \pi^0)
=
\frac{1}{\sqrt 2}\mathcal{M}_{\text{CS}, PV}(\Xi_b^-\to\Lambda_b \pi^-)
\end{align}

\item DPE
\begin{align}
\mathcal{M}_{\text{DPE}, PV}(\Xi_b^0\to\Lambda_b \pi^0)=0
\end{align}

\end{itemize}

\end{appendix}

\end{document}